\documentclass[longauth]{aa}  

\usepackage{orcidlink}
\usepackage{amssymb}
\usepackage{graphicx}

\usepackage{color}
\usepackage{xcolor}
\usepackage{lastpage}
\usepackage{multirow}
\usepackage{soul}

\usepackage{float}
\usepackage{txfonts}

\usepackage[]{hyperref}
\usepackage{subcaption}
\usepackage[switch]{lineno}
\usepackage[normalem]{ulem}

\newcommand{\delete}[1]{}

\newcommand{\changed}[1]{#1}

\newcommand{\RALow}{148.5}
\newcommand{\RAHigh}{151.8}
\newcommand{\DECLow}{0.6}
\newcommand{\DECHigh}{3.8}

\newcommand{\speczFlagThreeOrFour}{49802}
\newcommand{\speczFlagThirteenOrFourteen}{1475}
\newcommand{\broadTotal}{1420}
\newcommand{\broadWithSpec}{1420}
\newcommand{\broadSpeczFrac}{100.0}
\newcommand{\marchesiTotal}{3707}
\newcommand{\marchesiWithSpec}{1821}
\newcommand{\marchesiSpeczFrac}{49.1}
\newcommand{\ztfTotal}{516}
\newcommand{\ztfWithSpec}{300}
\newcommand{\ztfSpeczFrac}{58.1}
\newcommand{\vlaTotal}{5435}
\newcommand{\vlaWithSpec}{1539}
\newcommand{\vlaSpeczFrac}{28.3}
\newcommand{\assefTotal}{898}
\newcommand{\assefWithSpec}{369}
\newcommand{\assefSpeczFrac}{41.1}
\newcommand{\quaiaTotal}{368}
\newcommand{\quaiaWithSpec}{169}
\newcommand{\quaiaSpeczFrac}{45.9}
\newcommand{\vstTotal}{227}
\newcommand{\vstWithSpec}{206}
\newcommand{\vstSpeczFrac}{90.7}

\newcommand{\anyAll}{10100}
\newcommand{\anySpecz}{3866}

\begin{document} 
   \title{Photometric redshifts for active galactic nuclei with LePHARE for the Vera C. Rubin Observatory}

   \author{R.~Shirley\orcidlink{0000-0002-1114-0135}\thanks{rshirley@mpe.mpg.de}\fnmsep
          \inst{1},
M.~Salvato\inst{1,2}, 
J.~Cohen-Tanugi\orcidlink{0000-0001-9022-4232}\inst{3}, 
O.~Ilbert\inst{4}, 
S.~Arnouts\inst{4}, 
R.~Ansari\inst{5} 
R.~Assef\inst{6}, 
M.~Banerji\inst{7}, 
A.~Bongiorno\inst{8}, 
W.~N.~Brandt\orcidlink{0000-0002-0167-2453}\inst{9,10,11}, 
J.~Buchner\inst{1}, 
J.~Comparat\inst{12}, 
D.~Ili\'{c}\orcidlink{0000-0002-1134-4015}\inst{13}, 
A.~Kova{\v c}evi\'{c}\orcidlink{0000-0001-5139-1978}\inst{13}, 
J.~Kubica\inst{14,15}, 
B.~Laloux\inst{1,16}, 
O.~Lynn\inst{14,15}, 
A.~Malz\inst{14}, 
L.~Marchetti\inst{17,18,19}, 
C.~Mazzucchelli\inst{6}, 
T.~Mkrtchyan\inst{6}, 
K.~Nandra\inst{1}, 
D.~Oldag\inst{15,20}, 
C.~Ricci\inst{21},
W.~Roster\inst{1}, 
E.~Saremi\inst{7},  
S.~Satheesh-Sheeba\orcidlink{https://orcid.org/0009-0003-0654-6805}\inst{22},
T.~Tasnim.~Ananna\inst{23}, 
M.~J.~Temple\orcidlink{0000-0001-8433-550X}\inst{24}, 
M.~Vaccari\orcidlink{0000-0002-6748-0577}\inst{17,18,19}, 
A.~Viitanen\inst{21,8,25}, 
C.~Wolf\inst{26}, 
Z.~Yu\inst{9}, 
T.~Zhang\inst{27}, 
the LSST Dark Energy Science Collaboration, and the LSST Active Galactic Nuclei Science Collaboration.
  }
  \titlerunning{LePHARE for Rubin}
    \authorrunning{Shirley et al.}

\institute{\centering Max-Planck-Institut f\"ur extraterrestrische Physik, Giessenbachstr. 1, 85748 Garching, Germany
\and Exzellenzcluster ORIGINS, Boltzmannstr. 2, D-85748 Garching, Germany 
\and Université Clermont-Auvergne, CNRS, LPCA, 63000 Clermont-Ferrand, France
\and Aix-Marseille Université, CNRS, CNES, LAM, Marseille, France
\and Université Paris-Saclay, Université Paris-Cité, CEA, CNRS, AIM, 91191, Gif-sur-Yvette, France
\and Núcleo de Astronomía de la Facultad de Ingeniería, Universidad Diego Portales, Av. Ejército Libertador 441, Santiago, Chile
\and School of Physics and Astronomy, University of Southampton, Highfield Campus, Southampton SO17 1BJ, UK
\and INAF-Osservatorio Astronomico di Roma, Via Frascati 33, 00078 Monteporzio Catone, Italy
\and Department of Astronomy and Astrophysics, 525 Davey Lab, The Pennsylvania State University, University Park, PA 16802, USA
\and Institute for Gravitation and the Cosmos, The Pennsylvania State University, University Park, PA 16802, USA
\and Department of Physics, 104 Davey Laboratory, The Pennsylvania State University, University Park, PA 16802, USA
\and Université de Grenoble Alpes, CNRS, Grenoble INP, LPSC-IN2P3, 53, Avenue des Martyrs, 38000, Grenoble, France
\and University of Belgrade - Faculty of Mathematics, Department of Astronomy, Studentski trg 16, Belgrade, Serbia
\and McWilliams Center for Cosmology and Astrophysics, Department of Physics, Carnegie Mellon University, Pittsburgh, PA, USA
\and LSST Interdisciplinary Network for Collaboration and Computing Frameworks, 933 N. Cherry Avenue, Tucson, AZ 85721
\and INAF-Osservatorio Astronomico di Capodimonte, Via Moiariello 16, 80131 Napoli, Italy
\and Department of Astronomy, University of Cape Town, 7701 Rondebosch, Cape Town, South Africa
\and INAF $-$ Istituto di Radioastronomia, via Gobetti 101, 40129 Bologna, Italy
\and IDIA, Department of Astronomy, University of Cape Town, 7701 Rondebosch, Cape Town, South Africa
\and DIRAC Institute, University of Washington, Seattle, WA 98195, USA
\and Department of Astronomy, University of Geneva, ch. d’Ecogia 16, 1290, Versoix, Switzerland
\and Instituto de Astrofísica, Facultad de Ciencias Exactas, Universidad Andrés Bello, Fernández Concha 700, 7591538 Las Condes, Santiago, Chile
\and Department of Physics and Astronomy, Wayne State University, Detroit, MI 48202, USA
\and Centre for Extragalactic Astronomy, Department of Physics, Durham University, South Road, Durham DH1 3LE, UK
\and Department of Physics, University of Helsinki, PO Box 64, FI-00014 Helsinki, Finland
\and Research School of Astronomy and Astrophysics, Australian National University, Canberra, Australian Capital Territory, Australia
\and Department of Physics and Astronomy and PITT PACC, University of Pittsburgh, Pittsburgh, PA 15260, USA
 }
\date{Received 3 September 2025; accepted TBC}

   \date{Received 2026}

  \abstract
   {Active Galactic Nuclei (AGN) play a crucial role in galaxy evolution, but they are a minority of extragalactic sources with diverse Spectral Energy Distributions (SEDs), which depend on their means of selection. Upcoming large-scale surveys such as LSST will identify many AGN, but analysis tools are not optimized for them. The limited number of photometric bands in these surveys impacts the calculation of photometric redshifts for AGN, which are essential for scientific advancement.}
   {We use LePHARE to demonstrate the impact that a limited number of bands and erroneous assumptions have on the determination of the photometric redshifts of AGN. We conduct tests on six AGN samples selected using X-ray, radio, infrared, variability, color, and spectroscopic criteria in the COSMOS field, using photometry from HSC-CLAUDS, which is closest in depth and wavelength coverage to LSST.}
   {We present the LSST pipeline for LePHARE within the Redshift Assessment Infrastructure Layers (RAIL), facilitating comparison between SED fitting and machine learning algorithms. }
   {AGN that appear as point-like sources in optical data will be assigned highly unreliable photometric redshifts if they are processed using galaxy templates. Additionally, shallow all-sky surveys (like eROSITA, WISE, and ZTF) miss many AGN. As a result, these ``hidden'' AGN are often misidentified as galaxies in public survey data, leading to incorrect photometric redshifts.}
   {We provide the configurations that are suggested for each type of AGN alongside measures of expected performance as a function of redshift, magnitude, and selection. To facilitate studies with a panchromatic view of AGN, we also release photometric redshifts and posterior distributions for all AGN sources identified in the COSMOS field using the six criteria, based on 28-band photometry.}


   \maketitle

\section{Introduction}

It is now widely accepted that virtually every galaxy has a supermassive black hole at its center \citep[e.g.,][]{magorrian1998demography, Xiao_2011, Gultekin_2009, Gebhardt2000, Ferrarese2000}. At any given time, a fraction of these black holes are accreting large amounts of matter, which leads them to radiate energy as Active Galactic Nuclei (AGN). This contribution affects the Spectral Energy Distribution (SED) of a given source and therefore impacts both the estimates of the photometric redshift \citep[e.g., see Figure 4 of][]{salvato2019many} and the determination of the host's physical parameters when the contribution is not accounted for \citep[see discussion on XCigale and GRAHSP;] []{Yang2020, Buchner2024}.

This is the case for most of the catalogs of photometric redshifts and host physical parameters provided in the literature \citep[e.g.,][]{Laigle2016, Weaver_2022}.
At best, sources that appear as point-like in optical images are flagged or ignored, as they are assumed to be QSOs \citep[e.g.,][]{zhang2025, Comparat2017}, if not stars. This approach will eliminate the most obvious galaxies in the AGN phase, but will also remove very compact normal galaxies. At the same time, the majority of galaxies hosting an AGN (and thus with potentially incorrect redshift and physical parameter estimates) will remain unflagged unless additional methods for selecting AGN are adopted \citep[see for a review][]{Padovani2017}. It is possible to define samples of AGN, flag the sources for which photometric redshifts are not reliable in the available catalogs, and use techniques developed specifically for treating these objects. This can be done with both SED fitting (e.g., LePHARE \citep{Arnouts1999, ilbert2006}, EAZY \citep{brammer2008eazy}, X-Cigale \citep{Yang2020}) or machine learning (e.g., DNNZ \citep{nishizawa2020photometric}, CIRCLEZ \citep{saxena2024}) using bespoke AGN training sets.

For all-sky extragalactic surveys like the Legacy Survey of Space and Time \citep[LSST;][reaching a depth of $\sim 27.5$ magnitude in the $r$ band,]{ivezic2019lsst} and Euclid \citep[][reaching 24.0 magnitude (AB) in all three $YJH$ bands]{Mellier2025Euclid}, the situation is more complex. The depth of existing multi-wavelength data does not match that of LSST and Euclid. This means the SEDs are not well sampled, and many AGN will remain hidden in the main catalogs, affecting the cosmological studies and hampering the science focused specifically on AGN. 

The first aim of this paper is to increase awareness of the impact of AGN by quantifying the accuracy and the fraction of outliers for different AGN samples selected at various wavelengths, when photometric redshifts are computed either via SED fitting or machine learning, assuming they are galaxies. Not all AGN will be problematic, and it is important to know when estimates are reliable. AGN photometric redshifts have higher uncertainties than those computed for normal galaxies due to the unknown relative contribution of host and nuclear luminosity \citep[e.g.,][for an explanatory simulation]{hickox2018}. This makes it difficult to select a representative sample of templates. In particular, unobscured AGN are characterized by a negative-index power-law continuum SED, which is degenerate in redshift and luminosity and broad emission lines may be present.
 
The second aim of this paper is to, given a sample of AGN, provide a {\it vade mecum} for computing photometric redshifts reliably. The work presented here builds upon and extends work spanning twenty years \citep[e.g.,][]{salvato2009photometric,salvato2011dissecting, hsu2014ecdfs,ananna2017agn,norris2019,brescia2019,Fotopoulou2012LH,duncan2018photometric}, in particular working on radio and X-ray selected AGN.
For both aims, we use the SED fitting code LePHARE\footnote{Version used here: \url{https://doi.org/10.5281/zenodo.14162574}.} \citep[][]{Arnouts1999,ilbert2006} which is historically one of the most reliable and versatile algorithms \citep[see the PHAT challenge;][]{Hildebrandt2010, dahlen2013critical} both for galaxies and AGN. However, what is shown here can be implemented in other SED fitting algorithms such as Bayesian Photometric Redshifts \citep[BPZ;][]{Benitez2000,benitez2004faint,coe2006galaxies}, the Zurich Extragalactic Bayesian Redshift Analyzer \citep[ZEBRA;][]{feldmann2006zurich}, Hyperz \citep{bolzonella2000photometric}, Easy and Accurate Redshifts from Yale \citep[EAZY;][]{brammer2008eazy} and Phosphoros \citep[Paltani et al. in prep]{paltani2024}.\\
In support of these aims, we here present the LSST photometric redshift pipeline for galaxies and AGN implemented in The Redshift Assessment Infrastructure Layers \citep[RAIL,][]{RAIL2025}. RAIL enables large-scale comparison of multiple redshift-estimation algorithms and was used to compute photometric redshifts for Rubin’s Data Preview 1 \citep{zhang2025}.
The pipeline is designed to provide predicted redshifts from all libraries, including both galaxies and AGN, for all objects, enabling selection of the best estimate for a given sample at a later time.
Photometric redshifts for various AGN samples are computed here using data from the HSC-CLAUDS survey \citep{Sawicki2019} for two reasons. The data are of a quality and depth that will become available from LSST, and include NIR data that will become available when observations from LSST and Euclid are joined.
COSMOS also has a high fraction of sources with available spectroscopic redshifts \citep{khostovan2025}. Finally, the area is sufficiently large and well observed to host several thousand AGN identified across various wavelengths.\\

Following this introduction, we describe the photometric datasets in Section~\ref{sec:data_sets}. We then describe the six AGN samples investigated in Section~\ref{sec:agn_samples}.
In Section~\ref{sec:the_pipeline}, we provide technical details on LePHARE and the RAIL pipelines.
In Section~\ref{sec:results}, we present the performance on the existing data and provide sample-specific recommendations on which redshift estimate to use. 
In Section~\ref{section:discussion}, we discuss the results in a more general context. In Section~\ref{sec:releases}, we discuss the resulting redshifts that we are publishing both from HSC-CLAUDS and from the 28-band COSMOS2020 photometry. The conclusions are presented in Section~\ref{sec:conclusions}. We assume a standard flat $\Lambda$CDM cosmology throughout, with values $H_0=70$~km/s/Mpc, $\Omega_m$=0.3, and $\Omega_\Lambda$=0.7. 
All magnitudes are expressed in the AB system unless otherwise stated \citep{oke1974absolute}, for which a flux $f_\nu$ in microjansky ($10^{-29}$ erg cm$^{-2}$ s$^{-1}$ Hz$^{-1}$) corresponds to $\textrm{AB}_\nu = 23.9 - 2.5 \log_{10} (f_\nu /\mu \textrm{Jy})$.

\section{Datasets}\label{sec:data_sets}

The two datasets that will be used most widely for wide-area extragalactic studies in the future are LSST from Rubin and Euclid. 
LSST began the ten-year survey operations in July 2026. It will provide, over a decade, progressively deeper data in the Wide Fast Deep (WFD) footprint, covering the entire extragalactic sky visible from El Pe\~n\'on peak of Cerro Pach\'on (about 19,600~deg$^2$) and in the Deep Drilling Fields (DDFs; about 60~deg$^2$). 

The Euclid wide survey started taking data in 2024 \citep{scaramella2022euclid}. This dataset will eventually provide near-infrared coverage\footnote{The Nancy Grace Roman Telescope will also provide important near-infrared photometry but over 2400~deg$^2$ medium tier \citep{zasowski2025roman}.} across a large area of the LSST extragalactic sky. 

To test photometric redshifts for AGN using real data, we use existing deep data over the COSMOS field footprint \citep{Scoville2007COSMOS}, which is also one of the deep drilling fields of LSST and a calibration field for Euclid. 
In particular, we use the HSC-CLAUDS survey \citep{desprez2023}. This is a combination of $u$~band photometry from the CFHT MegaCam instrument with $grizy$ photometry from the Hyper-Suprime Cam (HSC), which in truth covers a much larger area (${\sim}20$~deg$^2$).

\begin{figure*}
\sidecaption
\includegraphics[width=12cm]{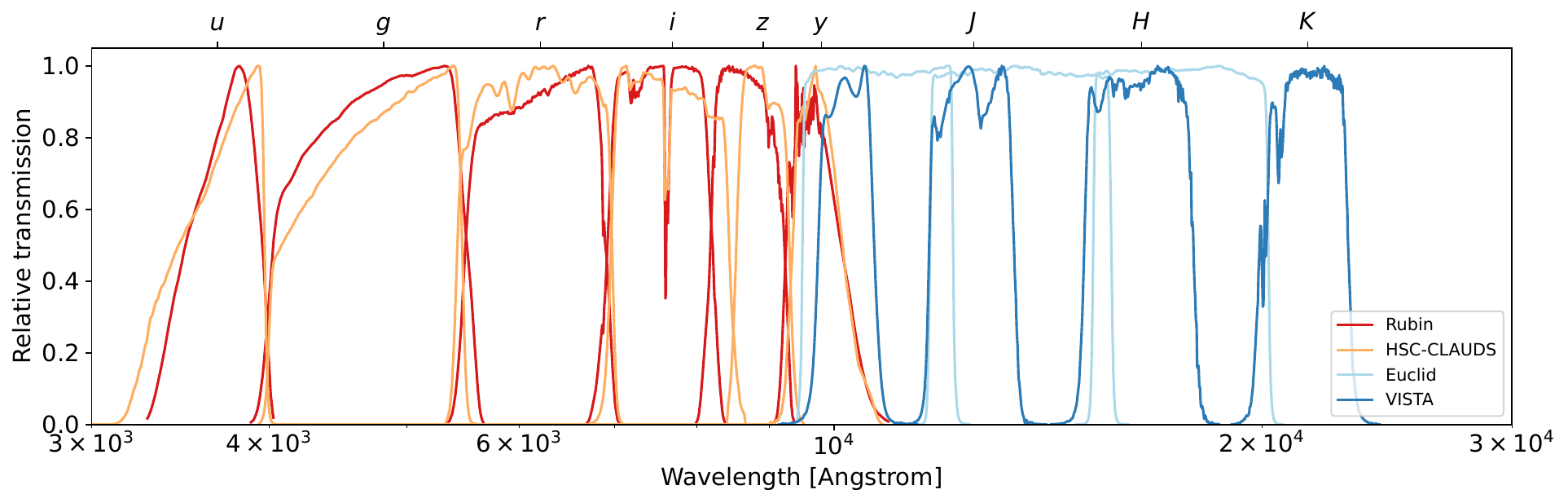}
\caption{All the filter transmission curves relevant to this work, normalised to 1. The LSST and HSC-CLAUDS filters are highly comparable. Euclid $J_E$ and $H_E$, and VISTA $J$ and $H$ filters are significantly different. 
}
\label{fig:filters_all}
\label{fig:filters}
\label{fig:filters_nir}
\end{figure*}

Figure~\ref{fig:filters} shows that the filter transmission curves for HSC-CLAUDS and LSST cover a similar region of wavelength space with comparable efficiency. 
Anticipating the time when Rubin and Euclid will eventually be jointly processed, we compute the photometric redshifts for the AGN samples with and without NIR data, for comparison (see Section~\ref{sec:performance}). As NIR data, we use again the catalog of \cite{desprez2023}, where the HSC-CLAUDS data are matched in coordinates to the UltraVISTA data.

\begin{table*}[ht]
    \caption{Overview of relevant existing and upcoming datasets for the work presented in this paper.}
    \centering
    \begin{tabular}{llccr} 
        \hline
        \hline
      &  Survey & bands  &  $5\sigma$ point source depth [mag] & area [deg$^2$]\\
        \hline
 \multirow{2}{*}[0.5\baselineskip]{Future} &       LSST WFD 10-year \citep{Bianco_2022} & $ugrizy$ & $i~26.8$ (AB) & 19 600 \\
     &   Euclid Wide \citep{scaramella2022euclid}& VIS~$YJH$ & VIS~$26.2$; H~$24.5$ (AB) & 14 000\\ 
        \hline
       & HSC-CLAUDS deep \citep{Sawicki2019} & $ugrizy$ & $i~27.1$ (AB) & 20 \\ 
 Existing     &  UltraVISTA \citep{mccracken2012ultravista} & $YHJK_s$ &$H~23.9$ (Vega) &1.5\\
       & VISTA VHS \citep{mcmahon2013first} & $JHK_s$ &  $H~19.8$ (Vega) & 20 000\\ 
        \hline
    \end{tabular}

    \label{tab:survey_depths}
\end{table*}

As seen in Figure~\ref{fig:filters_nir}, the $J$ and $H$ filters of Euclid are broader than those of VISTA, thus partially losing the ability to resolve features (e.g., presence of emission lines). This is particularly relevant for unobscured AGN and QSO, where the power-law continuum already limits our ability to determine the correct photometric redshift. At the same time, the contiguity of the filters prevents key features from falling in the gaps and introducing degeneracy \citep[seee.g.,][]{Benitez2009}.

The photometric redshifts computed here will be compared with reliable redshifts from the spectroscopic compilation of redshifts in the COSMOS field curated over the years by the COSMOS collaboration and presented in \citep{khostovan2025}. This compilation\footnote{\url{https://github.com/cosmosastro/speczcompilation}} combines literature measurements—including SDSS, DESI, LAMOST, and 2DF—with COSMOS-team DEIMOS observations targeting faint sources.

For most of the latter data, a visual inspection has been performed, and a quality flag for the redshift has been assigned, following \citet{hasinger2018deimos} and \citet{Lilly2007}. When possible, the quality flag from the original catalog from literature has been translated. The flag system indicates the probability of the redshift to be correct (including the case -99 when the data have very low signal-to-noise), or degenerate (when only one emission line is present; see \cite{khostovan2025} for more details). 
We retain only redshifts with a correctness probability above 75\%: flags 3, 4, 13, and 14, where “10+” denotes sources with visually identified broad emission lines.\footnote{This flag is activated only for a subsample of sources that have been visually inspected. The lack of flags 13 and 14 does not guarantee that the source lacks broad emission lines. Most of the sources flagged as 13 or 14 are actually X-ray detected sources from \citet{marchesi2016chandra}, that were used as targets during the spectroscopic campaign.}

There are \speczFlagThreeOrFour\ sources with the flag 3 or 4 and \speczFlagThirteenOrFourteen\ sources with the flag set to 13 or 14. Because this work is concerned chiefly with measuring photometric redshift performance, we use the right ascension and declination limits of the spectroscopic sample to define our area of interest. This corresponds to right ascensions between \RALow\ and \RAHigh\ and declinations between \DECLow\ and \DECHigh\ in equatorial coordinates. This was done to maximize the number of AGN with spectroscopic redshifts, but it means the selection functions of each sample are not uniform across the area.

\section{AGN Samples}\label{sec:agn_samples}
In this section, we define the six AGN samples designed to provide a representative overview of various selection methods. In the future, deeper data will identify even more AGN, particularly via variability at different cadences with LSST. In order of decreasing size of the spectroscopic sample, the six samples are:

\begin{itemize}

    \item Chandra: an X-ray selected sample of \marchesiTotal\ objects from the Chandra COSMOS Legacy survey \citep{civano2016chandra,marchesi2016chandra}. Of these, \marchesiWithSpec\ (\marchesiSpeczFrac\%) have a reliable spectroscopic redshift.
    
    \item VLA: from the Very Large Array VLA-COSMOS 3 GHz Large Project \citep{smolvcic2017vla}, we select the sources that are detected in radio and in the optical bands with a signal to noise higher than 5 and have the ratio between the flux density at 3GHz and in the $r$ band larger than 10 and classify them as radio-loud. Of these \vlaTotal\ sources, \vlaWithSpec\ (\vlaSpeczFrac\%) have a secure spectroscopic redshift.
    
    \item Broad-line: A sample of \broadTotal\ broad-line AGN and quasars from visual inspection of the \cite{khostovan2025} COSMOS spectroscopic compilation. Of these, \broadWithSpec\  (\broadSpeczFrac\%)  have a reliable spectroscopic redshift.
   
    \item WISE: A sample of \assefTotal\ WISE selected AGN based on the R90 (reliability $> 90\%$) criteria from the Wide-field Infrared Survey Explorer’s AllWISE Data Release \cite{assef2018}. Of these, \assefWithSpec\ (\assefSpeczFrac\%) have a reliable spectroscopic redshift.
    
    \item Optical variability: A sample of \ztfTotal\ variability selected AGN from the Zwicky Transient Facility (ZTF) based on light curve classification \cite{nakoneczny2025qzo}. Of these, \ztfWithSpec\ (\ztfSpeczFrac\%) have a reliable spectroscopic redshift. In addition, we consider the sample of \vstTotal\ sources selected to be AGN via optical variability in the VST $r$ band \citep{Decicco2019vst}. The sources cover a different parameter space than ZTF, with deeper optical data and shorter cadence. \vstWithSpec\ (\vstSpeczFrac\%) sources have reliable spectroscopic redshifts.
    
    \item Quaia: A sample of \quaiaTotal\ objects classified as extragalactic in Gaia, detected in unWISE and with spectroscopic redshifts obtained via machine learning using low-resolution spectra from Gaia, trained on a subsample with spectroscopy from SDSS \citep{StoreyFisher2024}. \quaiaWithSpec\ (\quaiaSpeczFrac\%) of these have a reliable spectroscopic redshift. 
    
\end{itemize}

\begin{figure}
\centering
\includegraphics[width=.85\columnwidth]{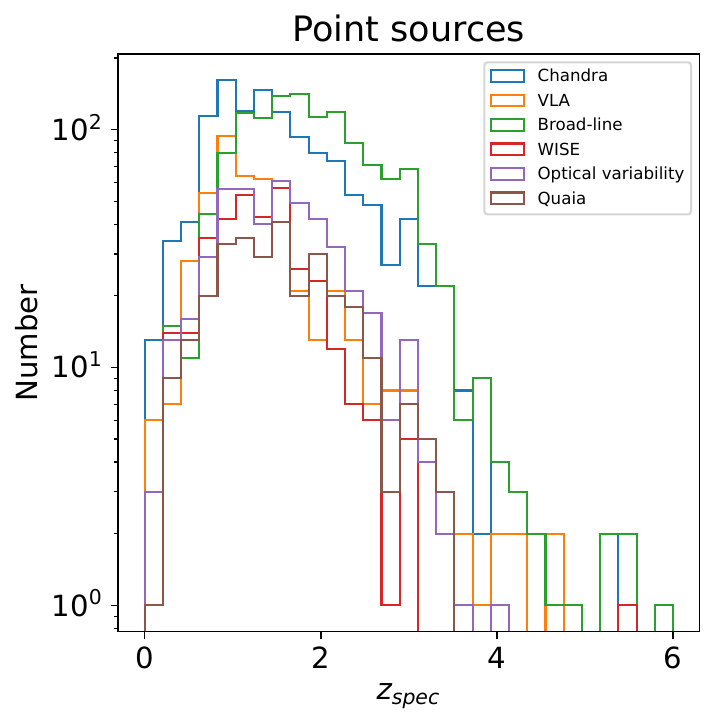}
\includegraphics[width=.85\columnwidth]{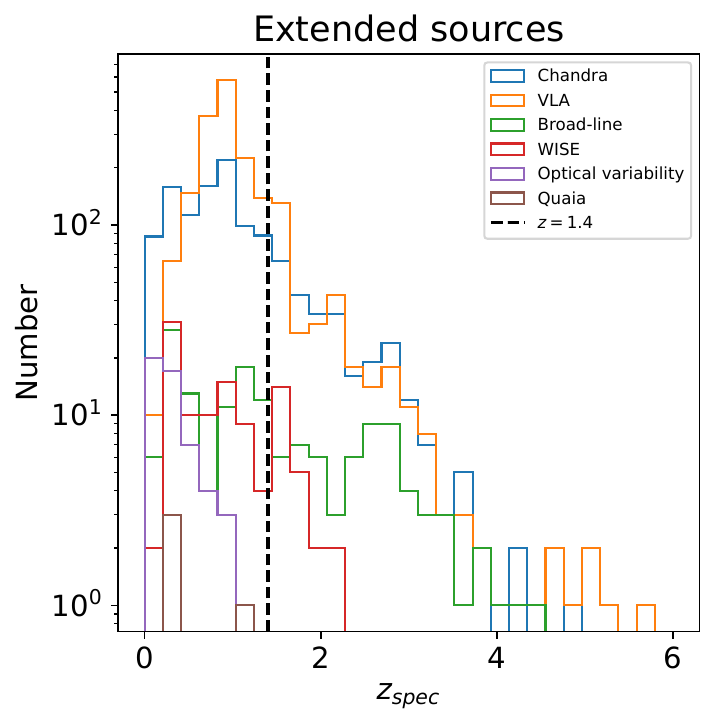}
\caption{Redshift distributions for the six AGN samples, divided by HSC $i$-band extendedness. Following Figure~\ref{fig:dimming}, the vertical dashed line marks the redshift beyond which typical isolated galaxies are unresolved in HSC, suggesting either blending or unreliable spectroscopic redshifts (see Section ~\ref{sec:performance}.)}
\label{fig:redshift_dist}
\end{figure}

\begin{figure*}
\centering
\includegraphics[width=1.\textwidth]{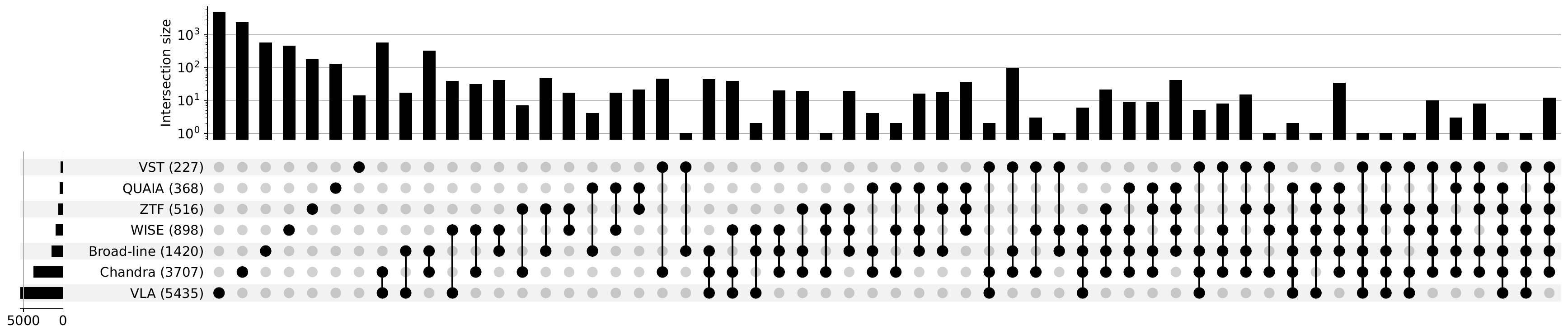}
\caption{The occupation of the various AGN samples and their intersections. No objects out of either the \anyAll\ objects in the total sample or the \anySpecz\ objects in the spectroscopic sample are in all the samples. Counts include only sources within the spectroscopic sample’s coordinate limits and with HSC-CLAUDS counterparts; spectroscopic subsample sizes are given in Section ~\ref{sec:agn_samples}.}
\label{fig:agn_intersections}
\end{figure*}

These six samples represent a wide range of AGN science cases and have different distributions of redshift (see Figure \ref{fig:redshift_dist}), black hole properties, and host properties \citep{zatarin2025,Padovani2017}. There are a total of \anyAll\ objects in the sample and \anySpecz\ with a spectroscopic redshift. Despite being all classified as AGN, the intersection of the samples is minimal (Figure~\ref{fig:agn_intersections}), and as we will show, this is reflected in the different recipes that are needed to compute reliable photometric redshifts \citep[see also][]{duncan2018agn}.
Even in a field like COSMOS, which has provided spectroscopic targets for more than 20 years, including for 10-meter-class telescopes, the sample of AGN with available spectroscopic redshifts remains highly incomplete, motivating the need for reliable recipes to compute photometric redshifts.

\section{The LSST photometric redshift pipeline}
\label{sec:optimising}\label{sec:the_pipeline}\label{sec:code_updates}
\begin{figure*}
\centering
\includegraphics[width=1.\textwidth]{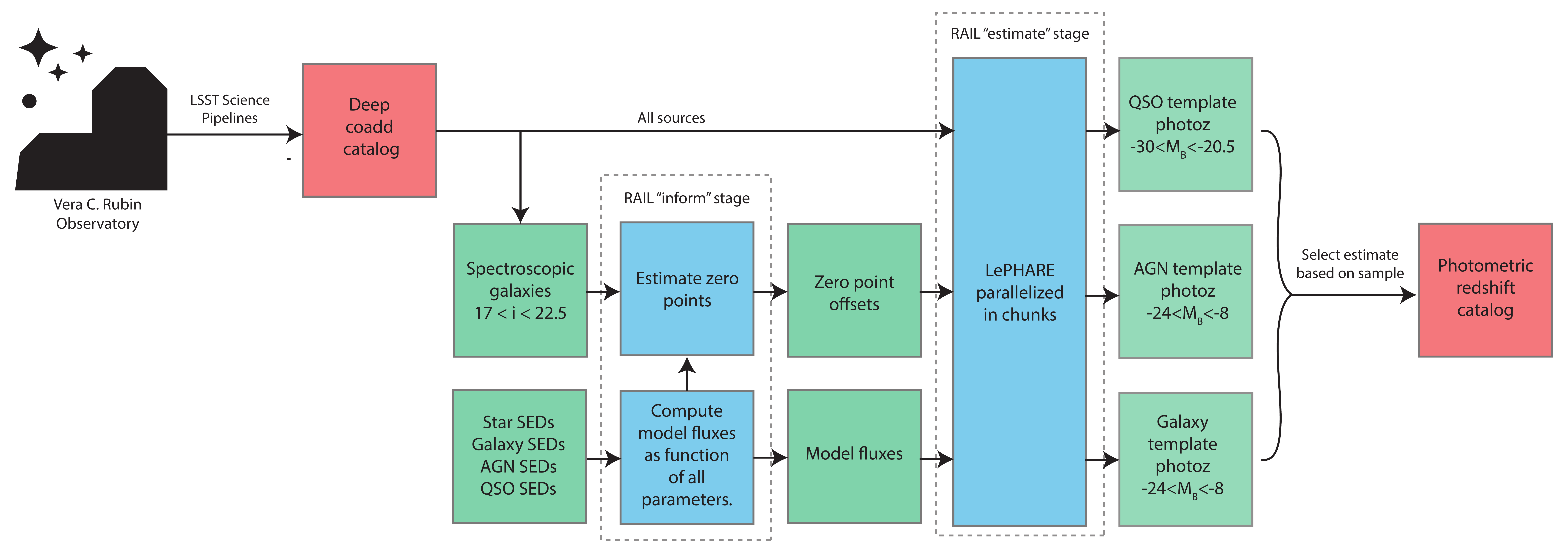}
\caption{Flow chart for large-scale LePHARE execution via RAIL, using separate quasar and AGN priors. Independent chunks enable straightforward parallelization; the estimate stage dominates CPU usage and scales linearly with source count. Final template and prior selection are described in the text}.
\label{fig:flowchart}
\end{figure*}

The Redshift Assessment Infrastructure Layers (RAIL) is a flexible open-source software library \footnote{\url{https://github.com/LSSTDESC/rail}} providing tools to produce at-scale photometric redshift data products from a number of algorithms. We integrated LePHARE into RAIL to deploy the pipeline for LSST. 
RAIL-LePHARE\footnote{\url{https://github.com/LSSTDESC/rail_lephare}}  will be used for future releases from LSST and has already been applied to Data Preview 1 (DP1) data \citep{zhang2025}. 
In this section, we report on recent changes to LePHARE in general and its integration into RAIL.
The overall pipeline is shown in Figure~\ref{fig:flowchart}.

LePHARE\footnote{\url{https://doi.org/10.5281/zenodo.14162574}}, which can estimate photometric redshifts and physical parameters, began as Fortran code \citet{Arnouts1999} and has undergone several incarnations. Recently, it has been rewritten in C\texttt{++}, with a PyBind-based interface that allows access and use from Python.

The code has three independent modules for fitting stars, galaxies, and AGN. The AGN module previously lacked features such as a) corrections for the intergalactic medium and intrinsic attenuation, b) the possibility to use priors, and c) redshift posterior distributions were all missing. All the work done in the past 20 years in computing photometric redshifts for AGN with LePHARE \citep[e.g.,][]{salvato2009photometric,salvato2011dissecting,Fotopoulou2012LH,hsu2014ecdfs,marchesi2016chandra,Nandra2015Aegis,ananna2017agn,Ni2021XMMS}, was done by using the ``galaxy'' module with some ad-hoc solutions tuned for AGN. The new code includes a complete AGN module. Given a sample of sources, the three modules are run independently, providing useful information on the quality of the fit by comparing results from the three libraries.

\subsection{LePHARE settings for the analysis}
\label{sec:lephare_settings}

Various studies over the last 15 years have shown that the optimal template library is determined by the depth of the X-ray data used for selection \citep[see, for example, Figure 2 of][for an update]{ananna2017agn}. Deep data from pencil-beam surveys require templates for galaxy-dominated AGN, as the brightest AGN will be missing from the survey. On the contrary, shallow data from wide areas will be rich in AGN-dominated sources, reaching also high redshifts. That is why, for every X-ray sample, different libraries have been assembled and tuned for that specific survey \citep{salvato2009photometric,salvato2011dissecting,Fotopoulou2012LH,hsu2014ecdfs,ananna2017agn,salvato2022erosita}. 

We adopt libraries used in previous work\footnote{We attempted to build a new library using an agnostic approach. We combined all available templates from various works and downsampled them to the best-fitting template at the spectroscopic redshift for each AGN type selected for this work. The new library created this way did not improve on the others already created.}. For the galaxy module, we used the library of \citet{ilbert2009cosmos}. For the AGN module, we tested the performances of two libraries: one, identified as \textit{QSO}, created for determining the photometric redshifts of the XMM-COSMOS sources \citep[][]{brusa2010xmmcosmos} presented in \citet{salvato2009photometric} and the other, identified as \textit{AGN}, created for the photometric redshifts of the optically extended sources detected by eROSITA in the eFEDS area \citep{brunner2022}, presented in \citet{salvato2022erosita}. Details on which templates are included in each library are in the original papers. Here, it is sufficient to report that in the \textit{QSO} library, the templates are either from empirically observed AGN and quasars or hybrid templates created by combining a galaxy with an AGN, with the AGN component dominating the SED. In the \textit{AGN} library, the templates mostly come from the compilation of \citet{brown2019spectral}; they are also hybrid templates that combine empirical spectra of galaxies and AGN, with a broader range of host/AGN contributions. We apply a prior on the absolute $g$ magnitude, set to -24,-8 for \textit{GAL} and \textit{AGN}, and -30,-20.5 for \textit{QSO}.

In Appendix~\ref{appendix:configuration_override}, we report the full configuration of LePHARE for extinction laws, nebular emission lines, and related settings for \textit{GAL}, \textit{AGN}, and \textit{QSO}.

\subsection{Milky Way Reddening}
Before using a SED-fitting method like LePHARE, the source fluxes are typically `de-reddened' to account for absorption by the Milky Way, mostly at shorter wavelengths. Usually, this is done by taking the value from dust extinction maps at a given sky position, computing the extinction in each survey photometric filter, and applying it to the measurements. This is equivalent to assuming the source's SED is flat in that band. In reality, each source has a distinct SED and is therefore reddened to varying degrees at a given wavelength. 

In X-ray spectral analysis, this effect is routinely accounted for \citep[see][]{wilms2000absorption}. 
\cite{galametz2017} investigated the impact of SED-dependence on photometric redshift estimation of normal galaxies and made recommendations for applying it in a realistic setting. 

We detail the method in Appendix~\ref{appendix:galactic_extinction}, and Figure~\ref{fig:reddening} shows its impact on the LSST bands. Because the band-pass correction depends on the source SED, reddening corrections can become large at high redshift when the SED is very red within the B band. This effect was absent from \cite{galametz2017}, which considered only $z<2$. The method is therefore most important for high-$E(B-V)$ regions and high-redshift sources, particularly quasars.
 
In the COSMOS field, with a median $E(B-V)=0.017$, the difference in performances obtained using the classical method or the new method proposed by \cite{galametz2017} is minimal (accuracy and outlier fractions differing by less than 1\%). However, over the wide fields of LSST with $E(B-V)>0.1$, the model flux extinctions can vary by up to 0.1 mag as a result of this SED dependence. DESC\footnote{\url{https://lsstdesc.org/}} science from the WFD survey will be limited to areas with $E(B-V)<0.2$. In the baseline cadence, up to 40\% of the sky has $E(B-V)>0.1$ \citep{olsen2018}.

\begin{figure}
\centering
\includegraphics[width=.8\columnwidth]{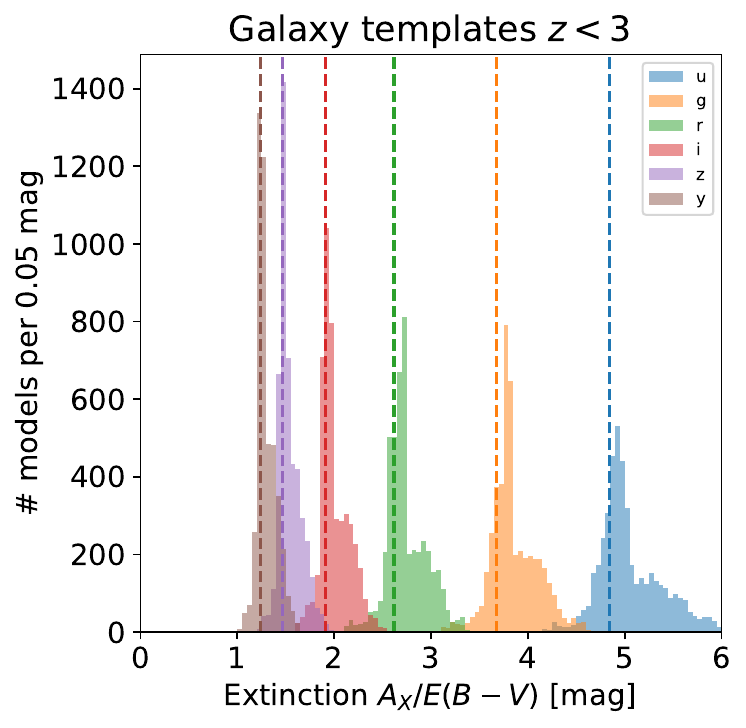}
\includegraphics[width=.8\columnwidth]{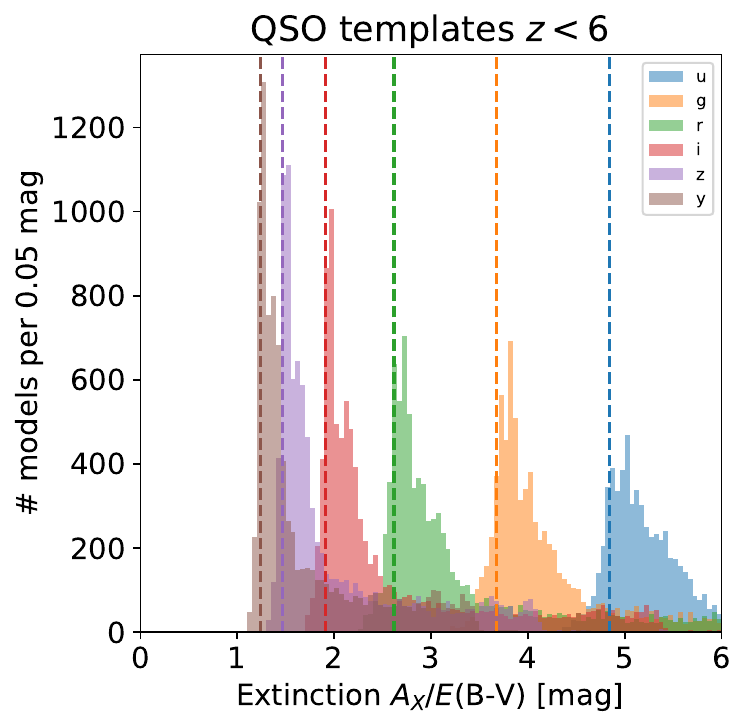}
\caption{Distribution of reddening for each model in the \texttt{GAL} library up to redshift $z=3$ (top panel) and for the \texttt{AGN} library up to redshift $z=6$ (lower panel). In both panels, the values are normalized by the ``$E(B-V)$'' value for B5 stars from the dust map, including the band-pass correction. Vertical lines show the singular value for a flat SED applied to all measurements using the traditional dereddening approach.}
\label{fig:reddening}
\label{fig:reddening_qsoz6}
\end{figure}

\section{Results}\label{sec:performance}\label{sec:results}

As commonly applied, the metrics used to assess the quality of the photometric redshifts are: bias, defined as
\begin{equation}
\beta =  {\rm median} \left( \frac{z_{\rm phot} - z_{\rm spec}}{1 + z_{spec}} \right)
\end{equation}
Standard deviation estimated from the normalized median absolute deviation:
\begin{equation}
\sigma_{\rm NMAD} = 1.48 \times {\rm median} \bigg|\frac{z_{\rm phot} - z_{\rm spec}}{1 + z_{\rm spec}} \bigg|
\label{eqn:nmad_biased}
\end{equation}
and fraction of outliers $\eta = n_{\rm outliers}/n_{\rm total}$ where the outliers are defined as 
\begin{equation}
\bigg|\frac{z_{\rm phot} - z_{\rm spec}}{1 + z_{\rm spec}}\bigg| > 0.15,
\end{equation}

We also compute the Probability Integral Transform \citep[PIT;][]{zhang2024} and Continuous Ranked Probability Score \citep[CRPS;][]{zhang2024}. These are statistical metrics used to evaluate the posterior probability distribution. While the CRPS assesses the accuracy and precision of a probabilistic forecast, the PIT evaluates the calibration of its probability density function.
They are defined as 

\begin{equation}
{\rm PIT} = \int_0^{z_{\rm spec}} p(z) \,\mathrm{d}z
\end{equation}
and 

\begin{equation}
   \text{CRPS}(F, z_{\rm spec}) = \int_{-\infty}^{\infty} 
    \left( F(z) - \mathbf{1}(z - z_{\rm spec}) \right)^2 dz
\end{equation}

\noindent respectively. $\mathbf{1}$ is the Heaviside step function, and $F$ is the cumulative distribution function for the posterior. In \cite{schmidt2020evaluation}, they produce a flat PIT distribution using the input spectroscopic redshift distribution as a posterior for all sources. This theoretically correct PIT distribution, even in the absence of predictive power, shows that it captures error performance regardless of point-estimate performance. The CRPS, in contrast, measures both accuracy and precision of the posterior, hence its frequent use as a loss function. We also compute the PIT outlier fraction, the fraction of objects with PIT values below 0.0001 or above 0.9999. This outlier fraction, which should total 0.0002 for an ideal uniform distribution, captures sources not represented in the template set that fail catastrophically and is analogous to the point-prediction outlier fraction.

AGN samples vary with wavelength, selection method, survey depth, and area. Wide, shallow surveys favor bright, low-redshift AGN, whereas deep surveys detect more faint and high-redshift sources.

Previous studies \citep[e.g.,][]{salvato2009photometric, Disanto18} show that morphology can break color–redshift degeneracies. AGN are classified as extended or point-like: extended sources are generally at low redshift, whereas point-like, AGN-dominated sources may lie at $z\geq6$. However, this classification is sensitive to resolution and seeing, which can cause misclassification \citep{hsu2014ecdfs}.

For this work, we split each AGN sample into extended and point-like sources using CLASS\_STAR\_HSC\_I = 0.75 as the separator. As seen in Figure~\ref{fig:redshift_dist}, in each AGN sample, there is a non-negligible fraction of sources that are classified as extended and are at a suspiciously high redshift. We demonstrate the point in Figure~\ref{fig:dimming} in Appendix~\ref{sec:appendix_extendedness}. There, we show at which redshift a galaxy with a diameter of 20 kpc, at the surface-brightness limit (26), with the HSC survey's pixel scale (0.168") and seeing (0.6"), cannot be resolved, as a function of its absolute magnitude. In order to quantify the impact that the extendedness classification has on the reliability of the the photometric redshift, we split the samples into low ($z<1.4$) and high ($z>1.4$) redshift bins, where $z=1.4$ is the redshift for which an isolated galaxy of $M_{\rm abs}$<-22 cannot be resolved in the HSC survey (see left bottom panel of Figure \ref{fig:dimming}).

For each AGN sample, the results are presented in an individual table, including metrics for each library (\textit{GAL/AGN/QSO}) computed with and without UltraVista photometry to demonstrate how the lack of NIR data will affect our results. In Figure~\ref{fig:stats_overview}, we show the key point prediction metrics for all the samples as a high-level overview. In table~\ref{tab:summary_table}, we highlight the best performing template set and prior for each sample. We note that the bias correlates with accuracy and approaches zero for the library that performs better for each AGN sample. Only in very few cases does it change significantly with the addition of NIR data to the optical. Below, we discuss the results for each AGN sample. 

\begin{table}
    \caption{Summary of best performing (i.e., smaller $\sigma_{NMAD}$) library per AGN sample, depending on morphology and number of bands available.}
    \centering
    \begin{tabular}{lcc}
        \hline
        \hline
Sample              & Point-like &  Extended \\
                    &  $ugrizy$/$ugrizyYJH$    &   $ugrizy$/$ugrizyYJH$        \\
        \hline
Chandra             & \textit{QSO/QSO} & \textit{GAL/GAL} \\
VLA                 & \textit{GAL/QSO} & \textit{GAL/GAL} \\
Broad-line          & \textit{QSO/QSO} & \textit{QSO/AGN} \\
WISE                & \textit{QSO/QSO} & \textit{GAL/GAL} \\
Optical variability & \textit{QSO/QSO} & \textit{GAL/GAL} \\
Quaia               & \textit{QSO/QSO} & \textit{AGN/QSO} \\

        \hline
    \end{tabular}

    \label{tab:summary_table}
\end{table}

Each AGN sample has a dedicated table reporting metrics for the \textit{GAL}, \textit{AGN}, and \textit{QSO} libraries, with and without UltraVista photometry. Figure ~\ref{fig:stats_overview} summarizes the key metrics, while Table ~\ref{tab:summary_table} identifies the best template set and prior for each sample. We note that the bias correlates with accuracy and approaches zero for the library that performs better for each AGN sample. The results for each sample are discussed below.

\subsection{Chandra Legacy COSMOS X-ray Sample}

Table~\ref{tab:marchesi_both} confirms the group’s results. For extended sources, \textit{GAL} performs best, achieving comparable low-redshift precision ($\sigma_{\rm NMAD}=0.06$) but reducing outliers from 16.9–21.1\% to 5.1\%. For point sources, \textit{QSO} performs best at all redshifts, including when NIR data are added. NIR data do not significantly improve precision but substantially reduce outliers. Although splitting extended sources by X-ray flux and photometric redshift is recommended \citep{salvato2011dissecting}, it is not applied here for simplicity.

\subsection{VLA Radio Sample}
Table~\ref{tab:vla_both} shows that for the radio-loud sources in COSMOS, the \textit{GAL} library performs the best, regardless of the extendedness of the sources. For the extended sources, the fraction of outliers is extremely low. The results are confirmed, and adding NIR data further implies that the photometric catalogs LSST will release are safe for these objects, unless the source is also classified as an AGN in other ways.

\subsection{Broad-line Sample}
One could argue that this sample does not need to be discussed, as by definition, a spectrum for visual inspection must already be available. However, it may be that only one broad line is identified in the spectrum, which would not be sufficient to reliably define the redshift, so that a photometric redshift would be needed. 

The selection also yields a pure Type-I AGN sample, allowing us to test whether the SED is AGN-dominated. Table ~\ref{tab:broad_both} shows that \textit{QSO} performs best for point sources at all redshifts, while \textit{AGN} performs slightly better for extended sources, consistent with their identification as local Seyferts.

\subsection{WISE Infrared Sample}
Given the area of COSMOS and the shallowness of the AllWISE data, the WISE-selected sample is small, and the statistical meaning of the result is limited (see Table~\ref{tab:assef_both}).  
For the extended sources, the differences in precision and the fraction of outliers are only marginal. This is consistent with the fact that this sample is only 90\% reliable and that a fraction of galaxies could also be star-forming, not necessarily AGN. For the point sources, the \textit{QSO} library performs the best.

\subsection{ZTF and VST Optical Variability Sample}
Table~\ref{tab:opt_both} shows that for this sample, the \textit{GAL} library works best for the extended sources, and the \textit{QSO} library for the point sources. 
\citet{simm2016pan} showed how the photometric redshift for optically varying AGN is affected by the photometry being collected over years. Thus, the photometric redshift for this type of AGN will be more reliable when it is computed using single passes of LSST data, as they will be sufficiently deep and the photometry in the various bands taken relatively close in time to not be affected by variability. The deep LSST coadds will be averaged over a long and consistent baseline unlike HSC which observed each bands over a small number of visits at different times.

\subsection{Quaia selected Sample}
We show the various performance values for the Quaia sample in Tables~\ref{tab:quaia_both}. 
The Quaia sample is created by combining Gaia sources (which are sensitive only to point-like or compact objects) that are also detected in the unWISE catalog \citep{schlafly2019unwise}. Thus, it is not surprising that in Table \ref{tab:quaia_both} there are only 4 sources classified as extended. For this sample, the \textit{QSO} library performs best, as expected.

\section{Discussion}
\label{section:discussion}
The previous section identified the best library for each AGN sample by morphological type.
Except for the ZTF and Quaia selections, many sources are both extended and at high spectroscopic redshift (Figures~\ref{fig:redshift_dist} and \ref{fig:dimming}), likely due to blending. The superior performance of the \textit{QSO} library for these sources suggests they are not genuinely extended. Thus, extendedness cuts do not fully prevent QSO contamination or catastrophic redshift errors\footnote{As demonstrated in \cite{hsu2014ecdfs}, 30\% of the sources at a depth of 24 magnitude in CDFS, have a different classification from ground and from space. This will also happen for the sources in common between LSST and Euclid}. Moreover, extended-emission detection depends on surface-brightness limits, seeing, and source profiles, as discussed in Appendix~\ref{sec:appendix_extendedness}.

\begin{figure*}
\centering

\begin{subfigure}{\textwidth}
    \centering
    \includegraphics[width=.85\textwidth]{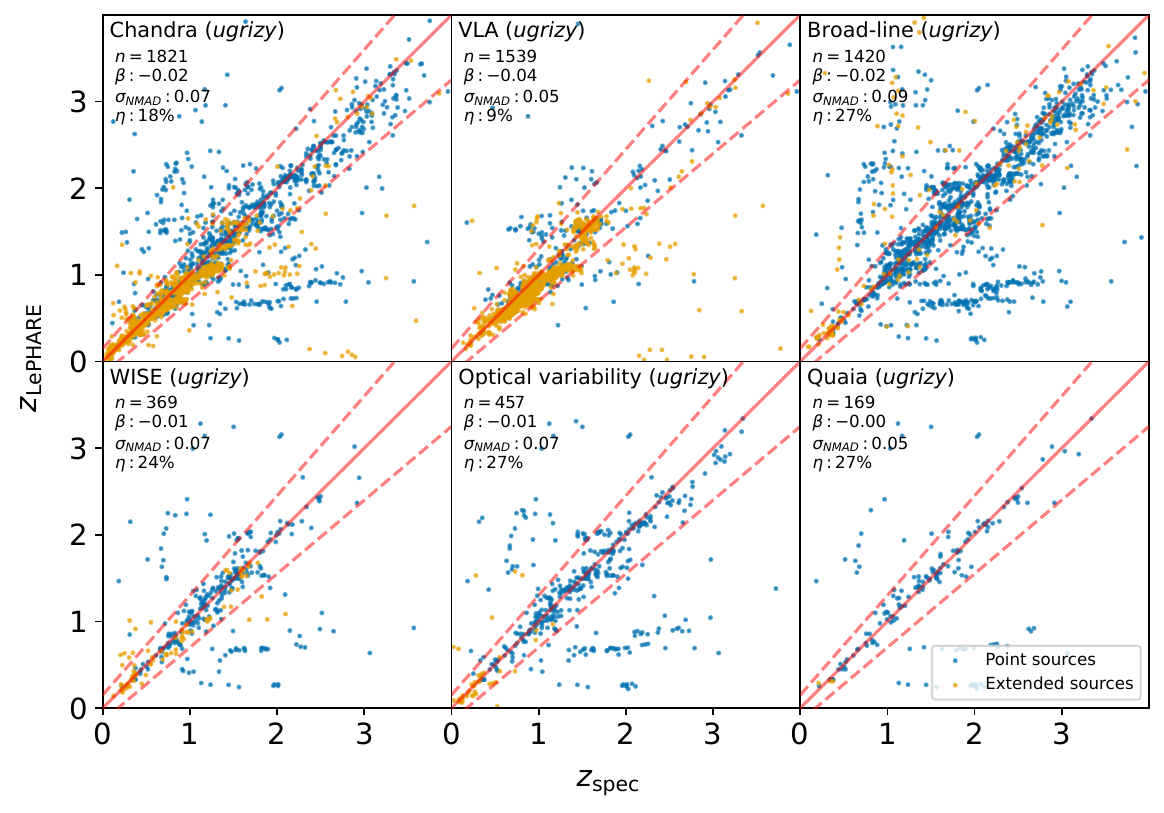}
\end{subfigure}

\vspace{0.5em}  

\begin{subfigure}{\textwidth}
    \centering
    \includegraphics[width=.85\textwidth]{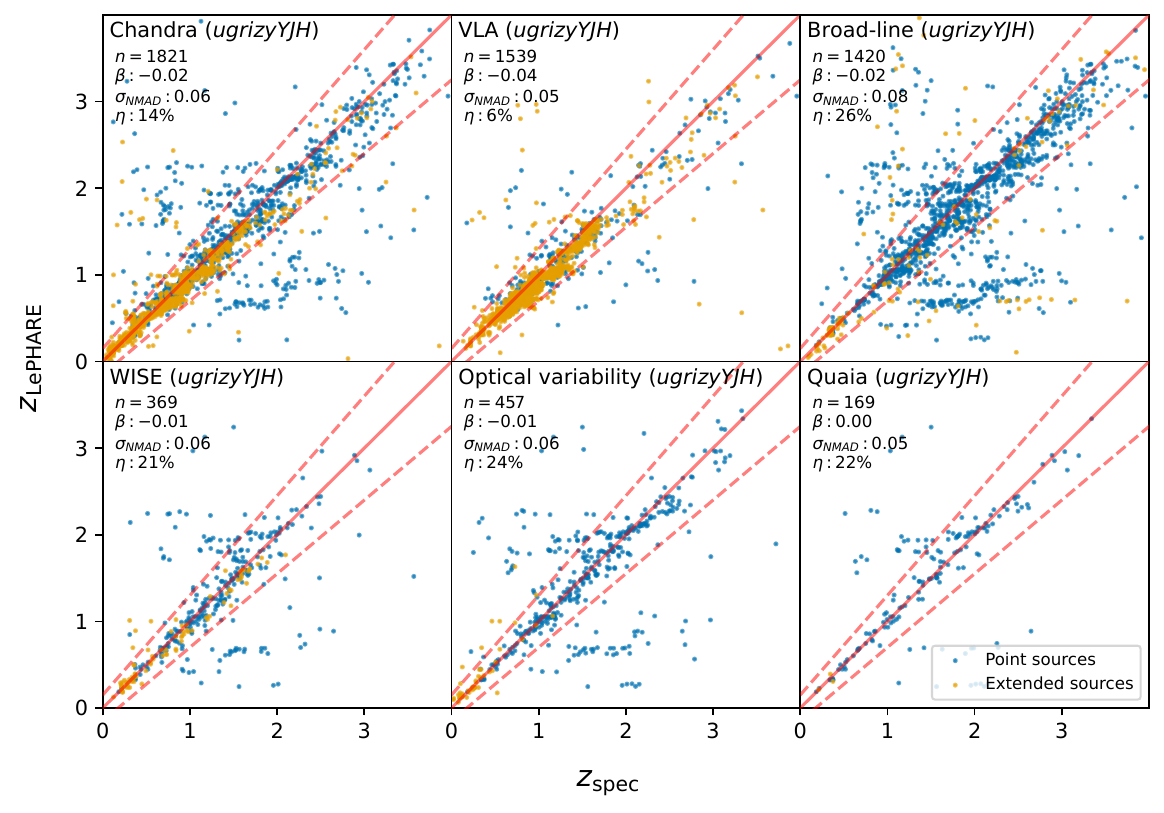}
\end{subfigure}

\caption{Photometric versus spectroscopic redshifts for the six AGN samples, using point estimates from the best-performing library. Point-like and extended sources are shown in blue and orange, respectively. The top and bottom panels use $ugrizy$ and $ugrizyYJH$ photometry, respectively. Solid lines indicate equality, while the regions beyond the dashed lines mark outliers.
}
\label{fig:zz_all_samples_both}
\end{figure*}

Looking at Figure~\ref{fig:zz_all_samples_both}, which visually summarizes the results from the various tables, we see that the point-like AGN in every panel suffer from outliers lying outside the dashed line and distributed along horizontal stripes. 
This is because they usually have SEDs dominated by a power law, with colors that are constant with redshift, and, with the limited photometry available, the degeneracy is not broken.

At first glance, the \textit{AGN} library may seem of limited use. This was the library that worked best for the computation of photometric redshifts for the X-ray detected sources in the wide area eROSITA/eFEDS \citep[140 square degrees;][]{brunner2022, salvato2022erosita} and Stripe 82-X\footnote{The \textit{AGN} library used in Stripe 82-X and eFEDS are very similar but not identical.} \citep[31 square degrees][]{ananna2017agn}. Wide-area surveys include numerous bright AGN at low redshift, which are hosted in extended galaxies, a population missing from the volume-limited COSMOS field.

COSMOS has deeper X-ray and radio data than the all-sky surveys, so many LSST-detected AGN will lack external AGN classifications and receive \textit{GAL}-based photometric redshifts. This is particularly problematic for X-ray sources. Table~\ref{tab:marchesi_both_xray_cut} repeats Table~\ref{tab:marchesi_both} considering only sources too faint for the eROSITA/DR2 \citep{ramos2026srg} flux limit ($F_{0.5-2{\text{ keV}}}<2.7 \times 10^{-14}{\text{ erg s$^{-1}$ cm$^{-2}$ }}$). Future machine-learning methods trained on AGN identified in deeper X-ray fields may enable AGN selection from LSST data alone.

One open question remains: whether the quality of the fitting using different libraries could be used to discriminate between galaxies and AGN. We have tried in many ways but unsuccessfully. However, thanks to advanced machine learning algorithms and the increasing size of the training samples, it would be possible to assign to each LSST source the probability to be a galaxy or an AGN \citep[e.g.,][]{Fotopoulou2018cpz,duncan2018photometric} or, at least, the probability to be an X-ray or radio emitter \citep[][]{salvato2022erosita,salvato2025counterpart, roster2024, roster2026FLASH}.

We tested two AGN samples selected via optical variability. Each LSST visit will be deeper than the visits of ZTF or VST-COSMOS. The cadence of the observations, however, will be comparable to that of ZTF (currently up to 8 years). This will provide new AGN samples, and we cannot yet predict which types of objects we will detect. Within the LSST AGNSC, there are already efforts to use variability as a parameter to measure redshift \citep[VAR-PZ][]{Satheesh-Sheeba2025}. Because of a lack of real training data, the method is currently not delivering competitive results. This may change in the future, and until then, the algorithm can provide useful priors for SED fitting.

\begin{figure}
\centering
\includegraphics[width=.8\columnwidth]{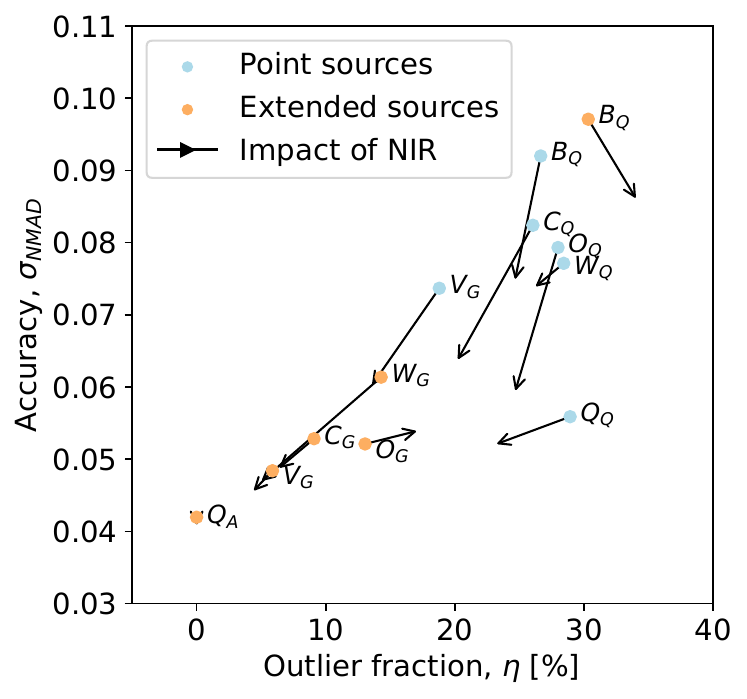}

\caption{The precision, $\sigma_{\mathrm{NMAD}}$, and outlier fraction, $\eta$, for the best-performing point estimate in each sample, with and without NIR data. Labels denote the sample’s initial and the optimal library (Gal, AGN, or QSO). Points show $ugrizy$ results; arrowheads show the corresponding $ugrizyYJH$ results.}
\label{fig:stats_overview}
\end{figure}

\subsection{The Impact of Near-Infrared Data}

\begin{figure}
\centering
\includegraphics[height=0.75\columnwidth]{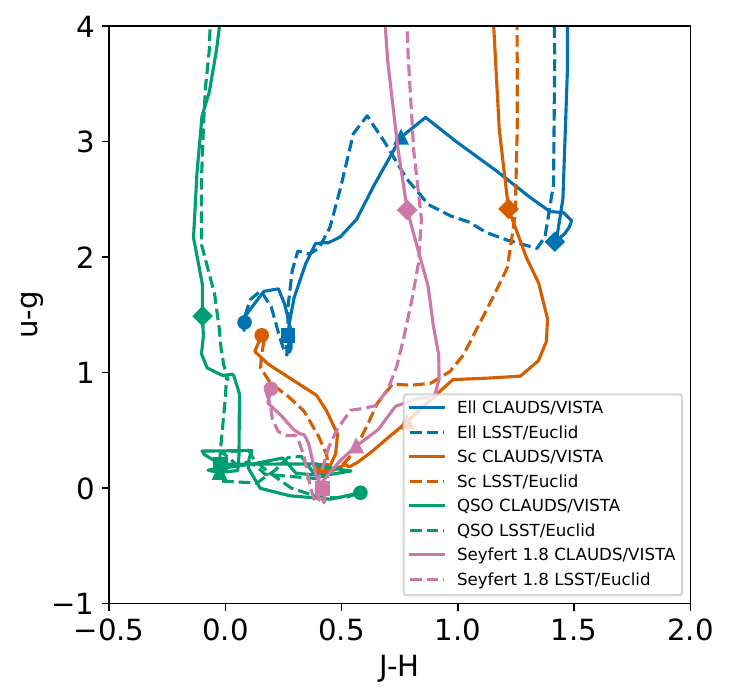}
\includegraphics[height=0.75\columnwidth]{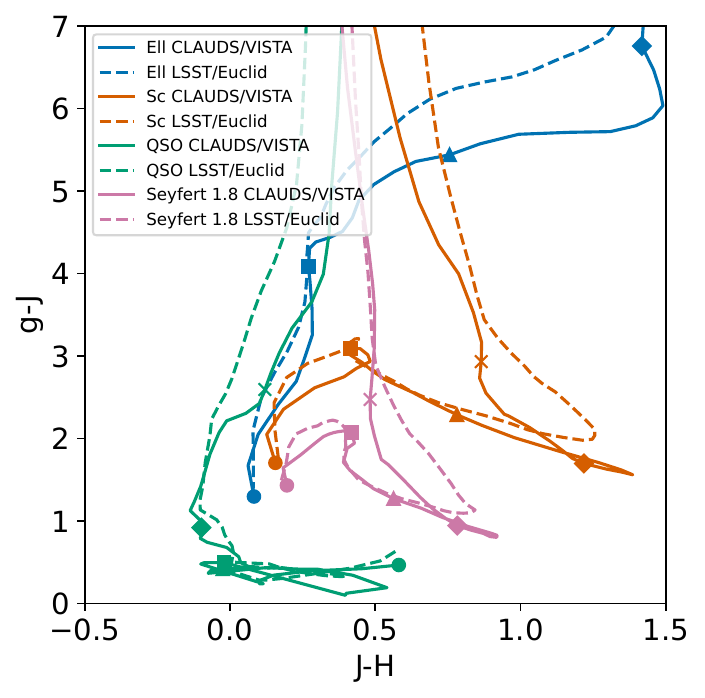}
\caption{Color-color plots as a function of redshift for an elliptical galaxy, a star-forming galaxy, a QSO, and a Seyfert 1.8 from the GAL, AGN and QSO libraries, using NIR data from Euclid (dashed lines) and Vista (solid lines). Markers are placed at redshifts 0~($\bullet$), 1~($\blacksquare$), 2~($\blacktriangle$), 3~($\blacklozenge$), and 4~($\times$).}
\label{fig:colours}
\end{figure}

For low-redshift extended AGN, the \textit{GAL} library achieves galaxy-like precision \citep[e.g.,][]{Weaver_2022}, whereas point sources show excessive outlier fractions at all redshifts due to photometric-band degeneracies \citep[see, e.g.,][]{salvato2009photometric}. Adding near-infrared data to the optical bands reduces outliers and improves accuracy across all samples (Figure~\ref{fig:stats_overview}).
Figure~\ref{fig:colours} shows the two sets of color-color tracks as redshift varies for optical and near-infrared colors using the existing datasets used in this paper and the upcoming filters from LSST and Euclid. While the tracks vary significantly for the elliptical galaxy, they still have recognizable features and contain comparable redshift information. In particular, for the QSO templates, the VISTA data offer a slightly better chance to break the degeneracy with fewer straight vertical lines. 

\subsection{Posterior metrics}\label{subsec:pit}

Table ~\ref{tab:crps_both} reports PIT outlier fractions and CRPS for point-like and extended sources in the six AGN samples as a function of the number of bands used. CRPS generally decreases with increasing band coverage, reaching its best values with 28 bands, except for extended sources in the Broad-line and optically varying samples. Unlike the increase in PIT outlier fraction reported for galaxies with additional bands in \cite{desprez2023}, the 28-band COSMOS data generally yield lower outlier fractions for AGN.
We show the PIT distribution in Figure~\ref{fig:overview} for the two largest AGN samples (X-ray and Radio), and for the library of templates that best fit point-like and extended sources, using $ugrizy$ photometry.

Figure~\ref{fig:crps} shows CRPS distributions for point-like and extended sources using six and nine bands, respectively. Adding NIR data shifts point-like sources toward lower CRPS values, while extended-source distributions remain nearly flat, indicating that the impact of NIR data depends on the AGN sample.

\begin{table*}[ht]

    \footnotesize
    \caption{\changed{Posterior distribution summary statistics for each of the six samples and three photometric data sets.}}\label{tab:crps_both}
        \centering
    \begin{tabular}{l  | ccc | ccc }
        \hline
        \hline
   &  \multicolumn{3}{c}{Point sources (6/9/30 band)}
& \multicolumn{3}{c}{Extended sources (6/9/30 band)} \\

\hline

      Sample & $n$ & PIT$_{\mathrm{out}}$ & $\overline{\mathrm{CRPS}}$
             & $n$ & PIT$_{\mathrm{out}}$ & $\overline{\mathrm{CRPS}}$   \\

        \hline

    Chandra  &  968/968/892
         &  0.50/0.59/0.43  
         &  0.33/0.29/0.22 

         &  853/853/785
         &  0.40/0.48/0.25  
         &  0.14/0.13/0.07 \\

    VLA  &  314/314/280
         &  0.57/0.68/0.45  
         &  0.23/0.21/0.17 

         &  1225/1225/1128
         &  0.52/0.62/0.18  
         &  0.15/0.15/0.07 \\

    Broad-line  &  1265/1265/935
         &  0.50/0.55/0.44  
         &  0.37/0.34/0.29 

         &  155/155/107
         &  0.53/0.62/0.49  
         &  0.54/0.50/0.42 \\

    WISE  &  285/285/221
         &  0.43/0.48/0.41  
         &  0.35/0.31/0.29 

         &  84/84/71
         &  0.46/0.54/0.20  
         &  0.13/0.10/0.07 \\

    Optical variability  &  411/411/346
         &  0.44/0.49/0.41  
         &  0.38/0.31/0.24 

         &  46/46/39
         &  0.30/0.33/0.33  
         &  0.08/0.09/0.08 \\

    Quaia  &  166/166/130
         &  0.36/0.44/0.42  
         &  0.39/0.30/0.26 

         &  3/3/3
         &  0.67/0.67/0.67  
         &  0.28/0.29/0.21 \\

    \end{tabular}
\end{table*}

\subsection{Comparison to Machine Learning}

The photometric redshifts planned for distribution by LSST\footnote{\url{https://dmtn-049.lsst.io/}} will be trained on extended sources \citep{zhang2025}. We test here how this will impact performance on the AGN samples. 
We test the results from two machine learning algorithms, $K$ Nearest Neighbour method \citep[KNN;][]{fix1985discriminatory} and FlexZBoost \citep{izbicki2017converting,Dalmasso2020}, by applying to our data the same procedure described in \cite{zhang2025}, who presented the photometric redshift performances using RAIL on the Vera C. Rubin Observatory Data Preview 1 (DP1).

We constructed a DP1-like training sample as in \cite{zhang2025} from extended sources with reliable redshifts (flag = 3 or 4), excluding known AGN, and added two-thirds of the extended sources from each AGN sample. The final sample contained 4811 sources with Galactic-extinction-corrected photometry. Results are shown in Tables~\ref{tab:knn} and~\ref{tab:fzboost}. Although both ML methods outperform SED fitting for extended AGN, they perform worse for point-like AGN. Unless LSST adopts more representative AGN training samples, its data releases may yield unreliable AGN redshifts. Until then, AGN studies should use SED fitting or bespoke, representative training sets.

\section{COSMOS2020 AGN photometric redshifts release}\label{sec:releases}

For the first time here, we assembled AGN selected by six methods in the same field and computed photometric redshifts from LSST- and Euclid-like data for sources with spectroscopic redshifts. Using the 28-band COSMOS2020 photometry \citep{weaver2023cosmos2020}, all metrics improve across samples (Figure ~\ref{fig:zz_all_samples_COSMOS_AGN}, the analog of Figure ~\ref{fig:zz_all_samples_both}), although the smaller survey area yields fewer sources than HST-CLAUDS. We provide a unified catalog containing the best available redshift (\texttt{Z\_BEST}), estimates from each library and band set, and PDZs, which will be available through VizieR and Zenodo upon publication.

\section{Conclusions}\label{sec:conclusions}

We present the pipeline for photometric redshifts for galaxies and AGN for LSST using the LePHARE SED fitting algorithm within the RAIL framework. In general, only six photometric bands are available, instead of the few tens available from pencil-beam surveys like e.g., COSMOS2020, MUSYC \cite{Gawiser2006MUSYC}, and SHARDS \cite{Perez2013shards}. 
We used the $ugrizy$ photometry from HSC-CLAUDS, which is very similar to LSST in depth and band coverage. 

We studied six COSMOS AGN samples, selected via X-ray, radio, MIR, optical/MIR, spectroscopy and optical variability. For sources with reliable spectroscopic redshifts from the compilation of \citet{khostovan2025}, we compared the photometric redshifts using three independent libraries and priors commonly used in the literature: \textit{GAL}, \textit{QSO}, and \textit{AGN}. The first includes only galaxies and has been widely used in COSMOS2020 and other data sets. The \textit{QSO} library includes AGN and AGN-dominant templates, the latter created by combining empirical templates of galaxies and AGN, as explained in \cite{salvato2009photometric}. It has been used for optically point-like AGN detected in various X-ray surveys. The \textit{AGN} library includes galaxies, AGN and hybrid sources and was produced for the optically extended AGN detected in wide areas like eROSITA/eFEDS and Stripe 82-X.

We find that AGN hosted in extended sources (and thus at low redshift; see Figure \ref{fig:redshift_dist} and \ref{fig:dimming}), the \textit{GAL} library performs best with precision of $\sim$0.09 but with the fraction of outliers depending on the AGN sample. However, previous works on AGN selected in wide-area surveys like Stripe 82-X \citet{ananna2017agn} and eROSITA/eFEDS \citet{salvato2022erosita} have shown that for AGN hosted in extended sources, the \textit{AGN} library is superior.
For most cases, the \textit{QSO} library performs well for the point-like AGN of all samples, with a precision of $\sim 0.08$ and a fraction of outliers around 20-25\%. Near-infrared photometric bands improve results, demonstrated here using the $YJH$ data from UltraVISTA, in lieu of Euclid or Roman.

In the COSMOS area, using the six criteria, we present a sample of $\sim10$k AGN, of which only 40\% have reliable spectroscopic redshifts. For the entire sample, we provide the photometric redshifts using the adequate library and the HSC-CLAUDS 9-band photometry and the full 28-band photometric data, where available. This sample provides a panchromatic view of a wide variety of AGN on the COSMOS field.

\begin{acknowledgements}
We thank the anonymous reviewer for constructive feedback, which helped significantly shape and hone the paper. This paper underwent internal review in the LSST Dark Energy Science Collaboration. Thank you to Sam Schmidt and Eric Charles for serving as the DESC publication review committee.

Based on observations collected at the European Southern Observatory under ESO programme ID 179.A-2005 and on data products produced by CALET and the Cambridge Astronomy Survey Unit on behalf of the UltraVISTA consortium.

We are grateful to the LSST Interdisciplinary Network for Collaborative Computing (LINCC) for funding and assistance. In particular, for the LINCC Frameworks Incubator `Developing the LePHARE photometric redshift code'.

JK, OL AM, DO, and TZ are supported by Schmidt Sciences. RA and SSS are supported by the ANID BASAL project.

This work made use of \texttt{Astropy}, a community-developed core \texttt{Python} package for Astronomy \cite{astropy:2013,astropy:2018,astropy:2022}.
We used Anthropic Claude Sonnet 4.6 for bug fixes, code improvements, and stylistic alterations to the text.

Contribution Statements:
R. Shirley: analysis, coordination, and development of rail\_lephare and LePHARE.
M. Salvato: analysis, coordination, and development of LePHARE AGN module.
J. Cohen-Tanugi, O. Ilbert, and S. Arnouts: core development of LePHARE and consulting on scientific performance.
D. Oldag, O. Lynn, and J. Kubica: code development consulting and work on the rail\_lephare and LePHARE packages.
All other authors contributed useful comments and edits.

\end{acknowledgements}

\section*{Data availability}
Hyper-Suprime-Cam Subaru Strategic Program Public Data are available at \url{https://hsc-release.mtk.nao.ac.jp}, while HSC-CLAUDS data are accessible at \url{https://www.clauds.net}. All other datasets referenced are public and available in the provided references. All data sets produced here will be made available following publication.

\appendix

\section{Milky Way extinction}\label{appendix:galactic_extinction}

\begin{figure}
\centering
\includegraphics[width=.8\columnwidth]{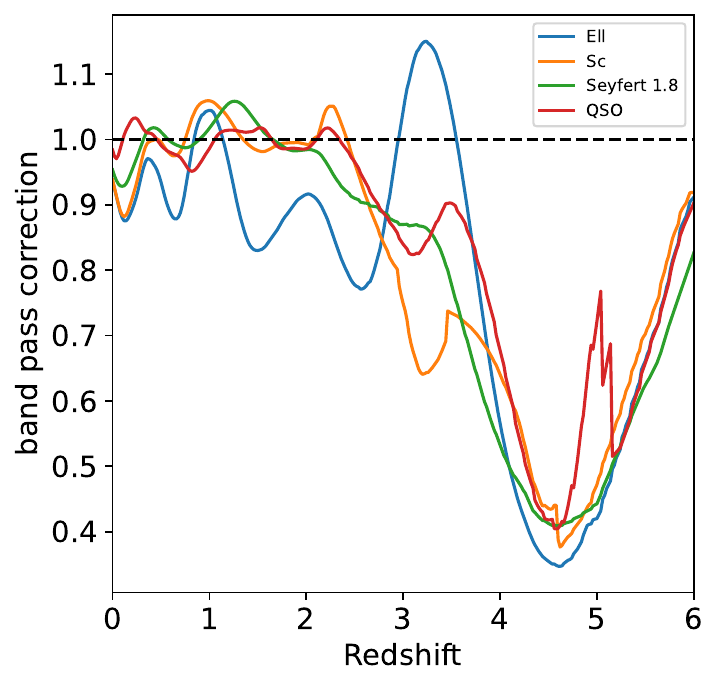}
\caption{Band-pass correction for a range of templates as a function of redshift. This factor scales the $E(B-V)$ values estimated for B5 stars from the Planck dust maps, calibrated using a sample of quasars. See main text for details.} 
\label{fig:extinction_g_redshift}
\end{figure}

A dust map such as those from \cite{schlegel1998maps} and \cite{schlafly2011measuring} provides the difference in the observed and intrinsic $B-V$ color for a particular SED, typically a B5 star. This $E^{B5}_{B-V}$ value is then used for correcting for extinction. The extinction then affects the source's full SED. Technically, one should compute the integrals over the models and filters for each particular value of $E^{B5}_{B-V}$, which would be prohibitively CPU-intensive.
\cite{galametz2017} instead compute the reddening in a given filter for a given model and assume a linear relation between the extinction and the dust column density \citep[see Figure A.1 of][]{galametz2017}.
We also compute the band-pass correction; the renormalization of the extinction law because the dust maps are calibrated assuming the B5-star SED rather than the SED of the source. This value, which is defined by the ratio of $E(B-V)$ for the source relative to a B5 star, is then used to correct the $E(B-V)$ value from the dust map. This follows directly from the definition of the band-pass correction given in Equation 6 of \cite{galametz2017} and the definition of $E^{SED}_{B-V}$ for a given source in Equation 3 of the same work. Figure~\ref{fig:extinction_g_redshift} shows the band-pass correction for a number of templates as a function of redshift for the $g$ band only. 

While for normal galaxies and galaxy-dominated AGN the range of band-pass corrections is within 20\% of unity up to redshift $z=2$, this is not the case at higher redshifts, where the band-pass correction goes as low as 0.4, meaning Milky Way extinction would be underestimated by a factor of 2.5. 
Overall, applying this method increases the computational time per object by 30~\% because it requires multiplying all model fluxes by the reddening factor. This can be avoided by binning objects in $E(B-V)$ such that it is only computed once for all objects in the bin.

\section{LePHARE configuration parameters}\label{appendix:configuration_override}

We present the key configuration parameters defining default RAIL runs on the full billion-object catalog, including the star, galaxy, \textit{AGN}, and \textit{QSO} libraries. Table~\ref{tab:lephare_config} lists the common parameters for \textit{GAL} and \textit{QSO/AGN} overrides.

\begin{table}
\small
\caption{Recommended and used LePHARE settings.}
\centering
\begin{tabular}{l p{5cm}}
\hline
\hline
Key & Value \\
\hline
COSMOLOGY & 70, 0.3, 0.7 \\
EB\_V & 0., 0.05, 0.1, 0.15, 0.2, 0.25, 0.3, 0.35, 0.4, 0.5 \\
EM\_DISPERSION & 1. \\
EM\_LINES & EMP\_UV \\
ERR\_FACTOR & 1.5 \\
ERR\_SCALE & 0.02, 0.02, 0.02, 0.02, 0.02, 0.02 \\
EXTINC\_LAW & SMC\_prevot.dat, SB\_calzetti.dat, SB\_calzetti\_bump1.dat, SB\_calzetti\_bump2.dat \\
FILTER\_LIST & u\_new.pb, gHSC.pb, rHSC.pb, iHSC.pb, zHSC.pb, yHSC.pb \\
MABS\_METHOD & 1 \\
MAGTYPE & AB \\
MAG\_ABS & -24, -8 \\
MAG\_REF & 2 \\
MIN\_THRES & 0.02 \\
MOD\_EXTINC & 18, 26, 26, 33, 26, 33, 26, 33 \\
NZ\_PRIOR & 2, -1 \\
RM\_DISCREPANT\_BD & 500 \\
TRANS\_TYPE & 1 \\
Z\_INTERP & YES \\
Z\_METHOD & BEST \\
Z\_RANGE & 0., 99.99 \\
Z\_STEP & 0.02, 0., 6. \\
\hline
QSO specific overrides:\\
\hline
MOD\_EXTINC & 1, 30 \\
EXTINC\_LAW & SMC\_prevot.dat \\
EB\_V & 0., 0.1, 0.2, 0.3, 0.4 \\
EM\_LINES & NO \\
MAG\_ABS\_QSO & -30, -20.5 \\
\hline
AGN specific overrides:\\
\hline
MOD\_EXTINC & 1, 999 \\
EXTINC\_LAW & SMC\_prevot.dat \\
EB\_V & 0., 0.1, 0.2, 0.3, 0.4 \\
EM\_LINES & NO \\
MAG\_ABS\_QSO & -24, -8 \\
\hline
\end{tabular}

\label{tab:lephare_config}
\end{table}

\section{Detailed sample results}
Here we include the detailed results on each of the samples. Tables~\ref{tab:marchesi_both}~to~\ref{tab:quaia_both} show the results for the 6 samples, and Tables~\ref{tab:knn} and \ref{tab:fzboost} show the results for the machine learning methods. Detailed discussion of the tables is in the main body of the text.

\begin{table*}[ht]
    \centering
    \caption{\changed{Photometric redshift performance for AGN selected by X-ray detection in Chandra.}}\label{tab:marchesi_both}
    \begin{tabular}{l c | c c c c | c c c c }
        \hline
        \hline
       & & \multicolumn{4}{c}{Point-like sources ($ugrizy$/$ugrizyYJH$)} & \multicolumn{4}{|c}{Extended sources ($ugrizy$/$ugrizyYJH$)} \\
        \hline
      Library/prior  & Redshift  & Sources & $\beta$ & $\sigma_{NMAD}$ & $\eta$ [\%] & Sources 
      & $\beta$ & $\sigma_{NMAD}$ & $\eta$ [\%]   \\

        \hline

\multirow[c]{2}{*}{GAL}  
             & $z<1.4$ & 455 & -0.03/-0.02 & 0.13/0.13 & 33.4/36.6 
                       & 730 & -0.02/-0.02 & 0.05/0.05 & 5.2/5.1  \\
\multirow[c]{2}{*}{$-24\!<\!M_g\!<\!-8$}  
         & $z\geq1.4$ & 513 & -0.35/-0.34 & 0.15/0.23 & 72.1/65.9 
                      &  123 & -0.07/-0.05 & 0.10/0.05 & 31.7/12.2  \\
           & $z>0.002$ & 968 & -0.12/-0.09 & 0.32/0.31 & 53.9/52.1 
                       & 853 & -0.02/-0.02 & 0.05/0.05 & 9.1/6.1  \\

        \hline

\multirow[c]{2}{*}{AGN} 
        & $z<1.4$ & 455 & -0.03/-0.04 & 0.14/0.09 & 41.8/33.0 
                 &  730 & -0.03/-0.03 & 0.06/0.05 & 15.2/8.8  \\
\multirow[c]{2}{*}{$-24\!<\!M_g\!<\!-8$}  
    & $z\geq1.4$ & 513 & -0.15/-0.13 & 0.31/0.26 & 51.9/50.8 
                 &  123 & -0.08/-0.05 & 0.15/0.08 & 39.0/23.6  \\
      & $z>0.002$ & 968 & -0.06/-0.06 & 0.18/0.14 & 47.1/42.4 
                 &  853 & -0.03/-0.03 & 0.07/0.06 & 18.7/11.0  \\

        \hline

\multirow[c]{2}{*}{QSO} 
        & $z<1.4$ & 455 & 0.01/0.00 & 0.08/0.06 & 25.1/18.0 
                 &  730 & 0.00/0.00 & 0.07/0.06 & 21.9/15.3  \\
\multirow[c]{2}{*}{$-30\!<\!M_g\!<\!-22.5$}  
    & $z\geq1.4$ & 513 & -0.04/-0.03 & 0.09/0.07 & 26.9/21.8 
                 &  123 & -0.03/-0.03 & 0.08/0.05 & 18.7/10.6  \\
      & $z>0.002$ & 968 & -0.01/-0.02 & 0.08/0.06 & 26.0/20.0 
                 &  853 & 0.00/-0.00 & 0.07/0.06 & 21.5/14.7  \\

\hline
    \end{tabular}
\end{table*}

\begin{table*}[ht]
    \centering
    \caption{\changed{Photometric redshift performance for AGN selected by radio detection from VLA.}}\label{tab:vla_both}
    \begin{tabular}{l c | c c c c | c c c c }
        \hline
        \hline
       & & \multicolumn{4}{c}{Point-like sources ($ugrizy$/$ugrizyYJH$)} & \multicolumn{4}{|c}{Extended sources ($ugrizy$/$ugrizyYJH$)} \\
        \hline
      Library/prior  & Redshift  & Sources & $\beta$ & $\sigma_{NMAD}$ & $\eta$ [\%] & Sources 
      & $\beta$ & $\sigma_{NMAD}$ & $\eta$ [\%]   \\

        \hline

\multirow[c]{2}{*}{GAL}  
             & $z<1.4$ & 195 & -0.04/-0.05 & 0.06/0.06 & 9.2/8.7 
                       & 1033 & -0.04/-0.04 & 0.05/0.05 & 2.5/3.2  \\
\multirow[c]{2}{*}{$-24\!<\!M_g\!<\!-8$}  
         & $z\geq1.4$ & 119 & -0.06/-0.05 & 0.11/0.07 & 34.5/27.7 
                      &  192 & -0.04/-0.04 & 0.08/0.04 & 24.0/9.4  \\
           & $z>0.002$ & 314 & -0.05/-0.05 & 0.07/0.06 & 18.8/15.9 
                       & 1225 & -0.04/-0.04 & 0.05/0.05 & 5.9/4.2  \\

        \hline

\multirow[c]{2}{*}{AGN} 
        & $z<1.4$ & 195 & -0.04/-0.05 & 0.11/0.09 & 30.3/22.6 
                 &  1033 & -0.05/-0.05 & 0.07/0.06 & 13.6/10.1  \\
\multirow[c]{2}{*}{$-24\!<\!M_g\!<\!-8$}  
    & $z\geq1.4$ & 119 & -0.04/-0.05 & 0.11/0.09 & 32.8/26.9 
                 &  192 & -0.05/-0.06 & 0.13/0.08 & 34.9/19.3  \\
      & $z>0.002$ & 314 & -0.04/-0.05 & 0.10/0.08 & 31.2/24.2 
                 &  1225 & -0.05/-0.05 & 0.07/0.06 & 16.9/11.5  \\

        \hline

\multirow[c]{2}{*}{QSO} 
        & $z<1.4$ & 195 & -0.01/-0.02 & 0.08/0.07 & 23.6/11.3 
                 &  1033 & -0.02/-0.02 & 0.06/0.05 & 19.1/11.9  \\
\multirow[c]{2}{*}{$-30\!<\!M_g\!<\!-22.5$}  
    & $z\geq1.4$ & 119 & -0.03/-0.04 & 0.07/0.06 & 21.0/16.8 
                 &  192 & -0.03/-0.04 & 0.07/0.04 & 19.8/13.0  \\
      & $z>0.002$ & 314 & -0.02/-0.03 & 0.08/0.06 & 22.6/13.4 
                 &  1225 & -0.02/-0.03 & 0.06/0.05 & 19.2/12.1  \\

\hline
    \end{tabular}
\end{table*}

\begin{table*}[ht]
    \centering
    \caption{\changed{Photometric redshift performance for AGN selected by broad line identification in the spectra.}}\label{tab:broad_both}
    \begin{tabular}{l c | c c c c | c c c c }
        \hline
        \hline
       & & \multicolumn{4}{c}{Point-like sources ($ugrizy$/$ugrizyYJH$)} & \multicolumn{4}{|c}{Extended sources ($ugrizy$/$ugrizyYJH$)} \\
        \hline
      Library/prior  & Redshift  & Sources & $\beta$ & $\sigma_{NMAD}$ & $\eta$ [\%] & Sources 
      & $\beta$ & $\sigma_{NMAD}$ & $\eta$ [\%]   \\

        \hline

\multirow[c]{2}{*}{GAL}  
             & $z<1.4$ & 347 & -0.07/-0.05 & 0.37/0.37 & 61.5/65.2 
                       & 91 & 0.04/0.02 & 0.11/0.12 & 35.2/37.4  \\
\multirow[c]{2}{*}{$-24\!<\!M_g\!<\!-8$}  
         & $z\geq1.4$ & 918 & -0.35/-0.35 & 0.16/0.22 & 73.6/70.1 
                      &  64 & -0.10/-0.08 & 0.15/0.10 & 42.2/31.2  \\
           & $z>0.002$ & 1265 & -0.34/-0.33 & 0.21/0.25 & 70.3/68.7 
                       & 155 & -0.01/-0.01 & 0.14/0.13 & 38.1/34.8  \\

        \hline

\multirow[c]{2}{*}{AGN} 
        & $z<1.4$ & 347 & -0.04/-0.03 & 0.19/0.15 & 48.6/46.0 
                 &  91 & -0.01/-0.00 & 0.11/0.06 & 31.9/27.5  \\
\multirow[c]{2}{*}{$-24\!<\!M_g\!<\!-8$}  
    & $z\geq1.4$ & 918 & -0.15/-0.14 & 0.32/0.28 & 52.3/52.1 
                 &  64 & -0.07/-0.07 & 0.15/0.14 & 45.3/43.8  \\
      & $z>0.002$ & 1265 & -0.09/-0.08 & 0.27/0.24 & 51.3/50.4 
                 &  155 & -0.03/-0.02 & 0.13/0.09 & 37.4/34.2  \\

        \hline

\multirow[c]{2}{*}{QSO} 
        & $z<1.4$ & 347 & 0.01/0.00 & 0.09/0.08 & 28.5/27.7 
                 &  91 & 0.02/0.03 & 0.09/0.09 & 36.3/34.1  \\
\multirow[c]{2}{*}{$-30\!<\!M_g\!<\!-22.5$}  
    & $z\geq1.4$ & 918 & -0.04/-0.03 & 0.09/0.08 & 25.9/23.4 
                 &  64 & -0.03/-0.04 & 0.08/0.08 & 21.9/18.8  \\
      & $z>0.002$ & 1265 & -0.02/-0.02 & 0.09/0.07 & 26.6/24.6 
                 &  155 & 0.01/0.00 & 0.10/0.09 & 30.3/27.7  \\

\hline
    \end{tabular}
\end{table*}

\begin{table*}[ht]
    \centering
    \caption{\changed{Photometric redshift performance for AGN selected by mid infrared detection in WISE.}}\label{tab:assef_both}
    \begin{tabular}{l c | c c c c | c c c c }
        \hline
        \hline
       & & \multicolumn{4}{c}{Point-like sources ($ugrizy$/$ugrizyYJH$)} & \multicolumn{4}{|c}{Extended sources ($ugrizy$/$ugrizyYJH$)} \\
        \hline
      Library/prior & Redshift  & Sources & $\beta$ & $\sigma_{NMAD}$ & $\eta$ [\%] & Sources 
      & $\beta$ & $\sigma_{NMAD}$ & $\eta$ [\%]   \\

        \hline

\multirow[c]{2}{*}{GAL}  
             & $z<1.4$ & 154 & -0.12/-0.13 & 0.32/0.32 & 58.4/61.1 
                       & 64 & -0.01/-0.02 & 0.06/0.05 & 12.5/6.2  \\
\multirow[c]{2}{*}{$-24\!<\!M_g\!<\!-8$}  
         & $z\geq1.4$ & 131 & -0.37/-0.37 & 0.09/0.11 & 96.0/91.3 
                      &  20 & -0.02/-0.03 & 0.05/0.02 & 20.0/0.0  \\
           & $z>0.002$ & 285 & -0.34/-0.34 & 0.16/0.19 & 75.5/75.0 
                       & 84 & -0.02/-0.02 & 0.06/0.05 & 14.3/4.8  \\

        \hline

\multirow[c]{2}{*}{AGN} 
        & $z<1.4$ & 154 & -0.09/-0.05 & 0.27/0.14 & 55.6/45.8 
                 &  64 & -0.03/-0.03 & 0.06/0.06 & 7.8/6.2  \\
\multirow[c]{2}{*}{$-24\!<\!M_g\!<\!-8$}  
    & $z\geq1.4$ & 131 & -0.49/-0.49 & 0.16/0.19 & 77.1/77.9 
                 &  20 & -0.06/-0.05 & 0.06/0.05 & 20.0/15.0  \\
      & $z>0.002$ & 285 & -0.35/-0.27 & 0.39/0.41 & 65.5/60.6 
                 &  84 & -0.03/-0.03 & 0.07/0.06 & 10.7/8.3  \\

        \hline

\multirow[c]{2}{*}{QSO} 
        & $z<1.4$ & 154 & 0.00/0.00 & 0.06/0.06 & 21.4/24.7 
                 &  64 & 0.01/0.01 & 0.06/0.06 & 23.4/23.4  \\
\multirow[c]{2}{*}{$-30\!<\!M_g\!<\!-22.5$}  
    & $z\geq1.4$ & 131 & -0.02/-0.01 & 0.12/0.09 & 36.6/27.5 
                 &  20 & -0.04/-0.03 & 0.10/0.04 & 25.0/5.0  \\
      & $z>0.002$ & 285 & -0.01/-0.00 & 0.08/0.07 & 28.4/26.0 
                 &  84 & -0.01/0.00 & 0.08/0.06 & 23.8/19.0  \\

\hline
    \end{tabular}
\end{table*}

\begin{table*}[ht]
    \centering
    \caption{\changed{Photometric redshift performance for AGN selected by optical variability in ZTF or VST.}}\label{tab:opt_both}
    \begin{tabular}{l c | c c c c | c c c c }
        \hline
        \hline
       & & \multicolumn{4}{c}{Point-like sources ($ugrizy$/$ugrizyYJH$)} & \multicolumn{4}{|c}{Extended sources ($ugrizy$/$ugrizyYJH$)} \\
        \hline
      Library/prior  & Redshift  & Sources & $\beta$ & $\sigma_{NMAD}$ & $\eta$ [\%] & Sources 
      & $\beta$ & $\sigma_{NMAD}$ & $\eta$ [\%]   \\

        \hline

\multirow[c]{2}{*}{GAL}  
             & $z<1.4$ & 166 & -0.02/-0.03 & 0.38/0.41 & 65.4/70.2 
                       & 46 & 0.02/0.01 & 0.05/0.05 & 13.0/17.4  \\
\multirow[c]{2}{*}{$-24\!<\!M_g\!<\!-8$}  
         & $z\geq1.4$ & 245 & -0.37/-0.38 & 0.10/0.12 & 93.7/91.7 
                      &  0 & -/- & -/- & -/-  \\
           & $z>0.002$ & 411 & -0.35/-0.36 & 0.16/0.18 & 82.2/83.0 
                       & 46 & 0.02/0.01 & 0.05/0.05 & 13.0/17.4  \\

        \hline

\multirow[c]{2}{*}{AGN} 
        & $z<1.4$ & 166 & -0.04/-0.04 & 0.20/0.12 & 50.0/46.1 
                 &  46 & -0.02/-0.04 & 0.08/0.06 & 21.7/10.9  \\
\multirow[c]{2}{*}{$-24\!<\!M_g\!<\!-8$}  
    & $z\geq1.4$ & 245 & -0.47/-0.48 & 0.19/0.22 & 70.2/71.3 
                 &  0 & -/- & -/- & -/-  \\
      & $z>0.002$ & 411 & -0.36/-0.34 & 0.40/0.40 & 62.0/61.1 
                 &  46 & -0.02/-0.04 & 0.08/0.06 & 21.7/10.9  \\

        \hline

\multirow[c]{2}{*}{QSO} 
        & $z<1.4$ & 166 & 0.01/0.00 & 0.07/0.06 & 22.9/26.5 
                 &  46 & 0.11/0.10 & 0.23/0.25 & 47.8/47.8  \\
\multirow[c]{2}{*}{$-30\!<\!M_g\!<\!-22.5$}  
    & $z\geq1.4$ & 245 & -0.03/-0.02 & 0.09/0.06 & 31.4/23.3 
                 &  0 & -/- & -/- & -/-  \\
      & $z>0.002$ & 411 & -0.01/-0.01 & 0.08/0.06 & 28.0/24.6 
                 &  46 & 0.11/0.10 & 0.23/0.25 & 47.8/47.8  \\

\hline
    \end{tabular}
\end{table*}

\begin{table*}[ht]
    \centering
    \caption{\changed{Photometric redshift performance for AGN selected by proper motion and color selected in Gaia and WISE (QUAIA).}}\label{tab:quaia_both}
    \begin{tabular}{l c | c c c c | c c c c }
        \hline
        \hline
       & & \multicolumn{4}{c}{Point-like sources ($ugrizy$/$ugrizyYJH$)} & \multicolumn{4}{|c}{Extended sources ($ugrizy$/$ugrizyYJH$)} \\
        \hline
      Library/prior  & Redshift  & Sources & $\beta$ & $\sigma_{NMAD}$ & $\eta$ [\%] & Sources 
      & $\beta$ & $\sigma_{NMAD}$ & $\eta$ [\%]   \\

        \hline

\multirow[c]{2}{*}{GAL}  
             & $z<1.4$ & 70 & -0.29/-0.31 & 0.18/0.16 & 76.9/81.2 
                       & 3 & 0.22/0.26 & 0.24/0.18 & 66.7/66.7  \\
\multirow[c]{2}{*}{$-24\!<\!M_g\!<\!-8$}  
         & $z\geq1.4$ & 96 & -0.39/-0.40 & 0.11/0.11 & 100.0/98.9 
                      &  0 & -/- & -/- & -/-  \\
           & $z>0.002$ & 166 & -0.37/-0.37 & 0.15/0.15 & 90.3/91.6 
                       & 3 & 0.22/0.26 & 0.24/0.18 & 66.7/66.7  \\

        \hline

\multirow[c]{2}{*}{AGN} 
        & $z<1.4$ & 70 & -0.07/-0.04 & 0.22/0.10 & 47.8/45.7 
                 &  3 & 0.01/0.00 & 0.04/0.05 & 0.0/0.0  \\
\multirow[c]{2}{*}{$-24\!<\!M_g\!<\!-8$}  
    & $z\geq1.4$ & 96 & -0.54/-0.54 & 0.09/0.10 & 95.8/95.8 
                 &  0 & -/- & -/- & -/-  \\
      & $z>0.002$ & 166 & -0.48/-0.48 & 0.22/0.24 & 75.8/74.5 
                 &  3 & 0.01/0.00 & 0.04/0.05 & 0.0/0.0  \\

        \hline

\multirow[c]{2}{*}{QSO} 
        & $z<1.4$ & 70 & 0.01/0.01 & 0.06/0.06 & 20.0/28.6 
                 &  3 & 0.02/0.02 & 0.06/0.04 & 0.0/0.0  \\
\multirow[c]{2}{*}{$-30\!<\!M_g\!<\!-22.5$}  
    & $z\geq1.4$ & 96 & -0.02/-0.01 & 0.08/0.05 & 35.4/18.8 
                 &  0 & -/- & -/- & -/-  \\
      & $z>0.002$ & 166 & -0.01/0.00 & 0.06/0.05 & 28.9/22.9 
                 &  3 & 0.02/0.02 & 0.06/0.04 & 0.0/0.0  \\

\hline
    \end{tabular}
\end{table*}

\begin{table*}[ht]
    \centering
    \caption{Photometric redshift performance for KNN on various AGN samples trained on a random 2/3 of all extended objects. }\label{tab:knn}
    \begin{tabular}{l c | c c c c | c c c c }
        \hline
        \hline
      &  & \multicolumn{4}{c}{Point sources} & \multicolumn{4}{|c}{Extended sources} \\
        \hline
     AGN sample  & Redshift & Sources  & $\beta$ & $\sigma_{NMAD}$ & $\eta$ [\%] & Sources & $\beta$ & $\sigma_{NMAD}$ & $\eta$ [\%]   \\

        \hline

\multirow{2}{*}[0.5\baselineskip]{Chandra}    & $z<1.4$ & 455 & -0.02 & 0.13 & 38.2 &  231 & -0.00 & 0.04  & 4.3  \\
        & $z\geq1.4$ & 513 & -0.37 & 0.27 & 77.0 &  42  & -0.02 & 0.05 & 21.4  \\

\hline

\multirow{2}{*}[0.5\baselineskip]{VLA}    & $z<1.4$ & 195 & -0.00 & 0.04 & 7.2 &  351 & -0.01 & 0.03  & 0.9  \\
        & $z\geq1.4$ & 119 & -0.08 & 0.14 & 41.2 &  51  & -0.01 & 0.05 & 13.7  \\

\hline

\multirow{2}{*}[0.5\baselineskip]{Broad-line}    & $z<1.4$ & 347 & -0.30 & 0.22 & 72.6 &  29 & -0.01 & 0.09  & 24.1  \\
        & $z\geq1.4$ & 918 & -0.38 & 0.28 & 77.9 &  26  & -0.09 & 0.20 & 46.2  \\

\hline

\multirow{2}{*}[0.5\baselineskip]{WISE}    & $z<1.4$ & 154 & -0.32 & 0.23 & 72.7 &  22 & -0.01 & 0.03  & 4.5  \\
        & $z\geq1.4$ & 131 & -0.54 & 0.15 & 97.7 &  7  & -0.02 & 0.03 & 14.3  \\

\hline

\multirow{2}{*}[0.5\baselineskip]{Optical variability}    & $z<1.4$ & 166 & -0.32 & 0.20 & 77.1 &  19 & -0.02 & 0.04  & 15.8  \\
        & $z\geq1.4$ & 245 & -0.54 & 0.20 & 97.1 &  0  & - & - & -  \\

\hline

\multirow{2}{*}[0.5\baselineskip]{Quaia}    & $z<1.4$ & 70 & -0.38 & 0.19 & 90.0 &  2 & -0.08 & 0.15  & 50.0  \\
        & $z\geq1.4$ & 96 & -0.57 & 0.15 & 100.0 &  0  & - & - & -  \\

\hline
    \end{tabular}
\end{table*}

\begin{table*}[ht]
    \centering
    \caption{\changed{same as Table ~\ref{tab:knn} but for FlexZBoost}}\label{tab:fzboost}
    \begin{tabular}{l c | c c c c | c c c c }
        \hline
        \hline
      &  & \multicolumn{4}{c}{Point sources} & \multicolumn{4}{|c}{Extended sources} \\
        \hline
     AGN sample  & Redshift & Sources  & $\beta$ & $\sigma_{NMAD}$ & $\eta$ [\%] & Sources & $\beta$ & $\sigma_{NMAD}$ & $\eta$ [\%]   \\

        \hline

\multirow{2}{*}[0.5\baselineskip]{Chandra}      & $z<1.4$ & 455 & -0.00 & 0.09 & 32.7 &  231 & 0.00 & 0.02  & 3.9  \\
        & $z\geq1.4$ & 513 & -0.33 & 0.16 & 79.7 &  42  & -0.02 & 0.05 & 21.4  \\

\hline

\multirow{2}{*}[0.5\baselineskip]{VLA}      & $z<1.4$ & 195 & -0.00 & 0.04 & 9.2 &  351 & -0.00 & 0.02  & 0.9  \\
        & $z\geq1.4$ & 119 & -0.13 & 0.23 & 47.1 &  51  & -0.01 & 0.04 & 13.7  \\

\hline

\multirow{2}{*}[0.5\baselineskip]{Broad-line}      & $z<1.4$ & 347 & -0.09 & 0.29 & 60.8 &  29 & 0.01 & 0.05  & 31.0  \\
        & $z\geq1.4$ & 918 & -0.32 & 0.18 & 78.5 &  26  & -0.17 & 0.24 & 61.5  \\

\hline

\multirow{2}{*}[0.5\baselineskip]{WISE}      & $z<1.4$ & 154 & -0.18 & 0.30 & 59.1 &  22 & 0.00 & 0.03  & 4.5  \\
        & $z\geq1.4$ & 131 & -0.37 & 0.14 & 94.7 &  7  & -0.07 & 0.05 & 14.3  \\

\hline

\multirow{2}{*}[0.5\baselineskip]{Optical variability}      & $z<1.4$ & 166 & -0.14 & 0.30 & 60.8 &  19 & 0.00 & 0.01  & 10.5  \\
        & $z\geq1.4$ & 245 & -0.38 & 0.14 & 94.7 &  0  & - & - & -  \\

\hline

\multirow{2}{*}[0.5\baselineskip]{Quaia}      & $z<1.4$ & 70 & -0.24 & 0.32 & 72.9 &  2 & 0.20 & 0.27  & 50.0  \\
        & $z\geq1.4$ & 96 & -0.42 & 0.18 & 96.9 &  0  & - & - & -  \\

\hline
    \end{tabular}
\end{table*}

\section{Extended vs. point-like}\label{sec:appendix_extendedness}

Throughout the paper, we used object extendedness as a prior for template selection, although it imperfectly distinguishes point-like AGN from extended sources. Its accuracy depends on image quality, pixel scale, source density, and—at fixed observational conditions—the source surface brightness and redshift.

\begin{figure*}
    \centering
    \includegraphics[width=\linewidth]{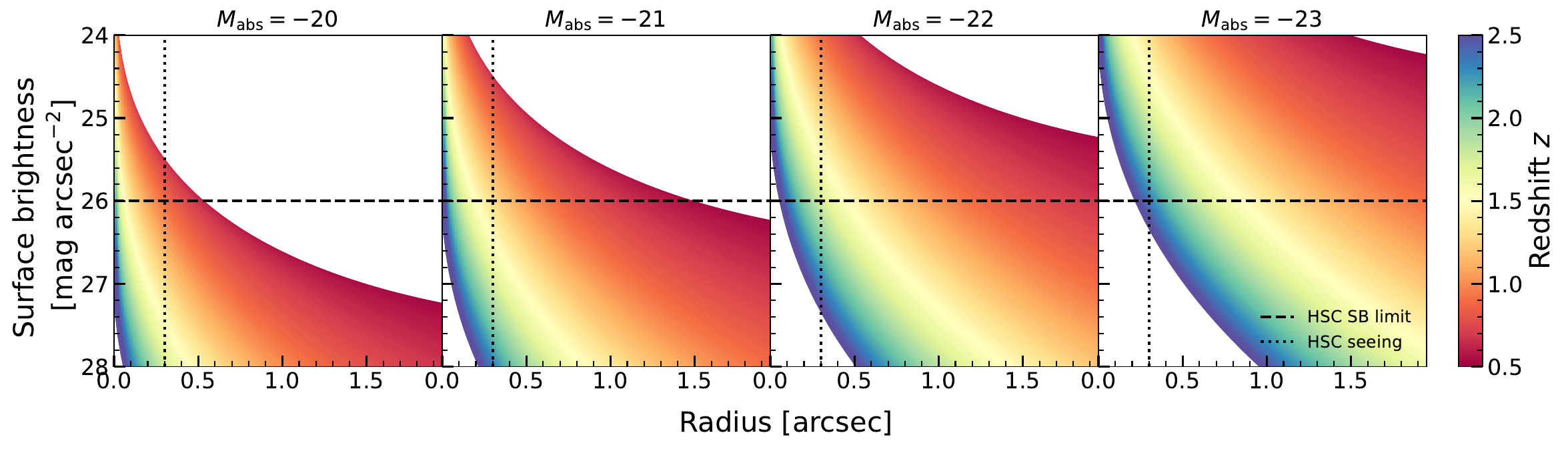}

\caption{S\'ersic surface-brightness profiles for inactive galaxies with $M\_{\rm abs}=-20$ to $-23$, $n=2.5$, and $R\_e=10$ kpc, illustrating the loss of detectable extended emission with redshift. The dashed line marks the HSC $i$-band detection limit, and the dotted line indicates its median seeing radius.}
    \label{fig:dimming}
\end{figure*}

Figure~\ref{fig:dimming} shows surface brightness versus radius across redshifts and absolute magnitudes, including the HSC $i$-band surface-brightness limit. Using typical seeing, it estimates the redshift beyond which host-galaxy extendedness becomes undetectable, explaining point-like classifications despite host contributions. Conversely, extended classifications for AGN-dominated sources may result from blending.

\begin{table*}[ht]
    \centering
    \caption{\changed{Photometric redshift performance for AGN detected by Chandra-COSMOS but too faint to be detected in eRASS:3.}}\label{tab:marchesi_both_xray_cut}
    \begin{tabular}{l c | c c c c | c c c c }
        \hline
        \hline
       & & \multicolumn{4}{c}{Point sources ($ugrizy$/$ugrizyYJH$)} & \multicolumn{4}{|c}{Extended sources ($ugrizy$/$ugrizyYJH$)} \\
        \hline
      Library & Redshift  & Sources & $\beta$ & $\sigma_{NMAD}$ & $\eta$ [\%] & Sources 
      & $\beta$ & $\sigma_{NMAD}$ & $\eta$ [\%]   \\

        \hline

\multirow{2}{*}[0.5\baselineskip]{Galaxies} 

    & $z<1.4$ & 427 & -0.02/-0.02 & 0.12/0.12 & 31.6/34.5 
              & 718 & -0.02/-0.02 & 0.05/0.05 & 4.6/4.6  \\
& $z\geq1.4$ & 503 & -0.35/-0.34 & 0.15/0.23 & 71.7/65.3 
             &  122 & -0.07/-0.05 & 0.10/0.05 & 32.0/12.3  \\
  & $z>0.002$ & 930 & -0.11/-0.09 & 0.32/0.30 & 53.2/51.1 
              & 840 & -0.02/-0.02 & 0.05/0.05 & 8.6/5.7  \\

        \hline
\multirow{2}{*}[0.5\baselineskip]{AGN}

    & $z<1.4$ & 427 & -0.03/-0.04 & 0.13/0.09 & 41.5/32.6 
             &  718 & -0.03/-0.03 & 0.06/0.05 & 15.3/8.8  \\
& $z\geq1.4$ & 503 & -0.14/-0.13 & 0.30/0.25 & 51.5/50.4 
             &  122 & -0.08/-0.05 & 0.16/0.09 & 39.3/23.8  \\
  & $z>0.002$ & 930 & -0.06/-0.06 & 0.18/0.14 & 46.9/42.2 
             &  840 & -0.04/-0.03 & 0.07/0.06 & 18.8/11.0  \\

        \hline
\multirow{2}{*}[0.5\baselineskip]{QSO} 

    & $z<1.4$ & 427 & 0.01/-0.00 & 0.08/0.06 & 25.3/17.3 
             &  718 & 0.00/0.00 & 0.07/0.06 & 21.7/15.0  \\
& $z\geq1.4$ & 503 & -0.04/-0.04 & 0.09/0.07 & 26.2/21.3 
             &  122 & -0.03/-0.03 & 0.08/0.05 & 18.9/10.7  \\
  & $z>0.002$ & 930 & -0.02/-0.02 & 0.08/0.06 & 25.8/19.5 
             &  840 & 0.00/-0.00 & 0.07/0.06 & 21.3/14.4  \\

\hline
    \end{tabular}
\end{table*}

\begin{figure*}
\centering

    \includegraphics[width=.9\textwidth]{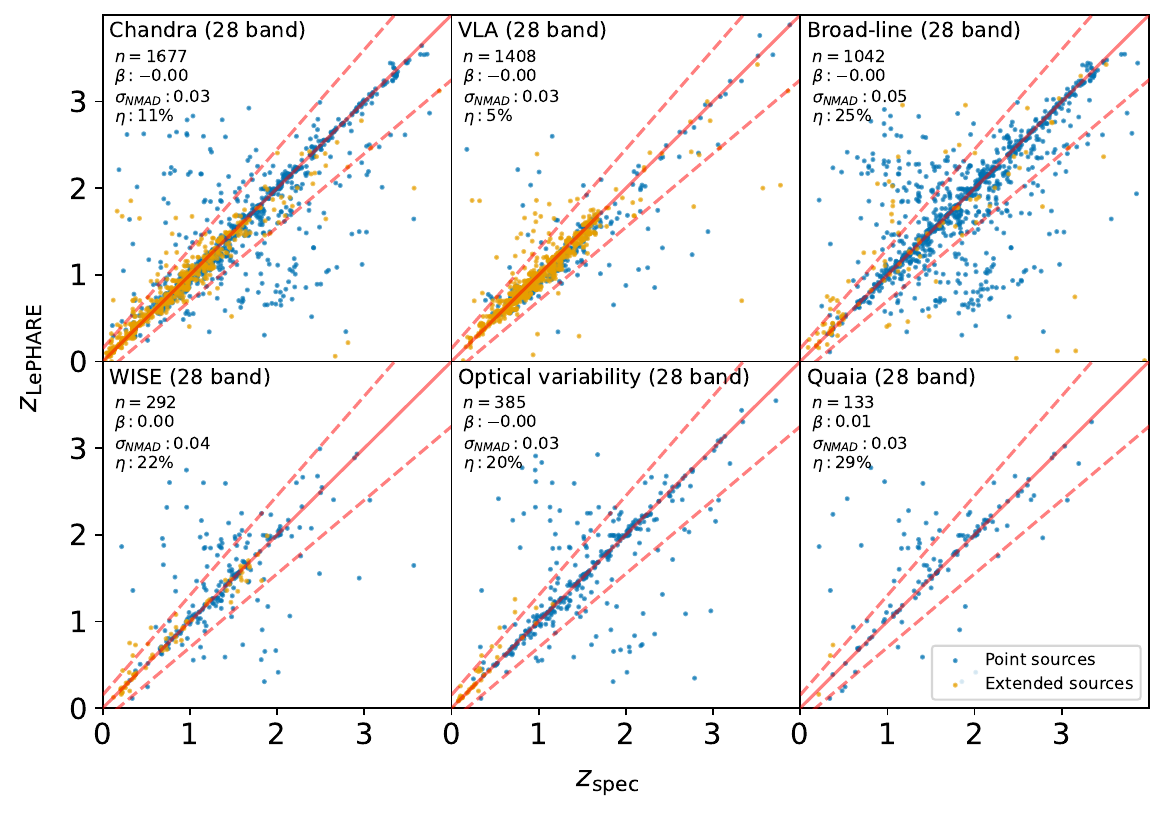}

\caption{Photometric versus spectroscopic redshifts for six AGN samples, using the lowest-$\sigma_{\mathrm{NMAD}}$ template library and combining point-like (blue) and extended (red) sources. Results use the 28-band \cite{Weaver_2022} CLASSIC photometry. }

\label{fig:zz_all_samples_COSMOS_AGN}
\end{figure*}

\begin{figure*}
    \centering
\includegraphics[width=1\textwidth]{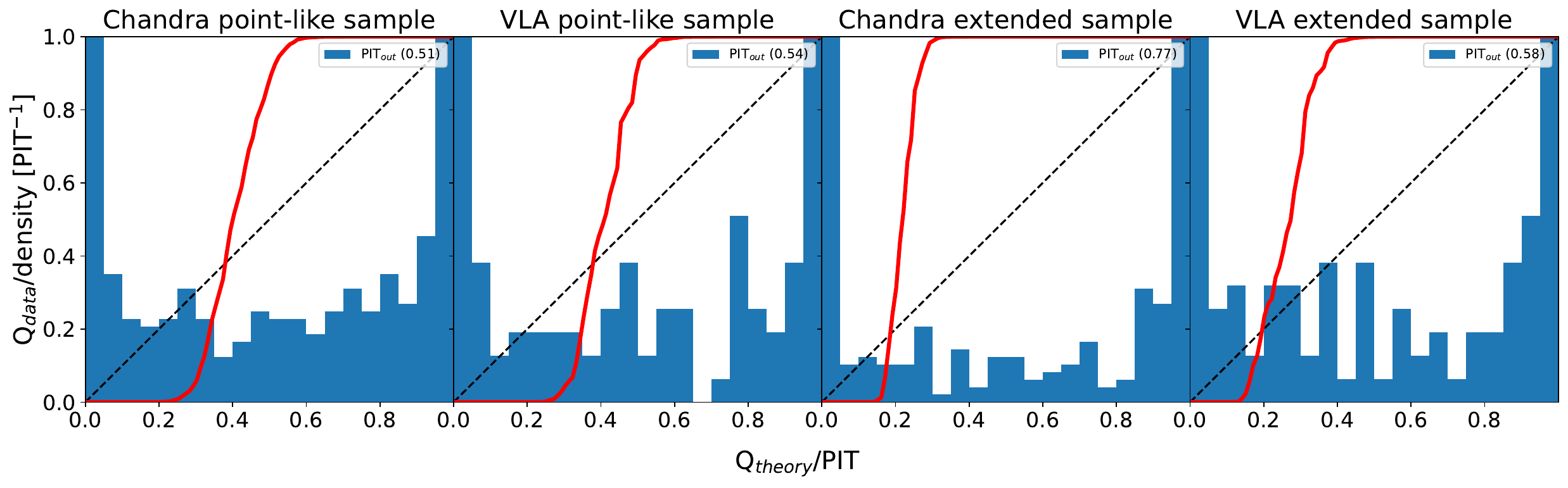}
    \caption{PIT distributions and quantile-quantile plots for the X-ray and radio sources (the largest of the samples) adopting the \textit{QSO} library for the point-like, and the \textit{GAL} library for the extended samples. The U shape indicates under-dispersion in the posteriors.}
    \label{fig:overview}
\end{figure*}

\begin{figure*}
    \centering
        \includegraphics[width=1\textwidth]{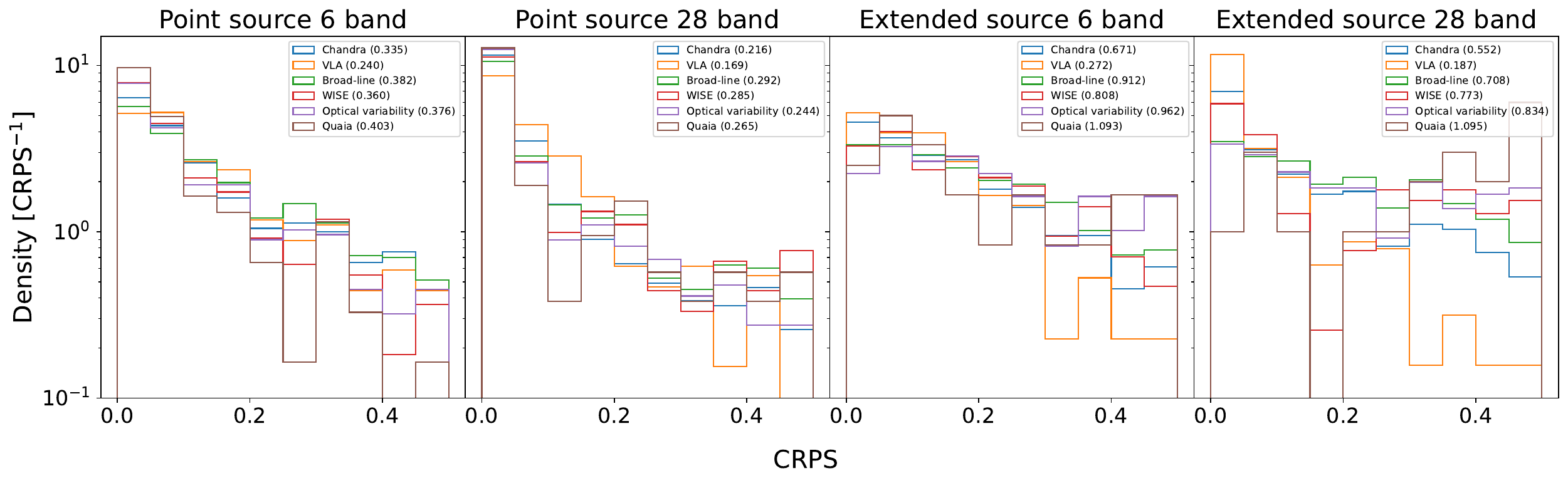}

    \caption{The distribution of CRPS values for point sources and extended sources based on 6-band and 28-band photometry. In general, the values are lower as more bands are included, as the estimates become more accurate.}
    \label{fig:crps}
\end{figure*}

\bibliographystyle{aa}
\bibliography{bibliography.bib}

\end{document}